\documentclass[10pt,journal,compsoc]{IEEEtran}

\usepackage{lineno,hyperref}
\usepackage{graphicx}
\usepackage{float}
\usepackage{enumitem}
\graphicspath{ {./images/} }
\usepackage[utf8]{inputenc}
\usepackage[english]{babel}
\usepackage{xcolor}
\usepackage{soul}
\usepackage{subfig}
\usepackage{amssymb}
\usepackage{amsmath}
\usepackage{algorithm}
\usepackage{algorithmicx}
\usepackage[noend]{algpseudocode}

\begin{document}

\title{Flow-Based Energy Services Composition}

\author{Amani~Abusafia,
        Abdallah~Lakhdari,
        and~Athman~Bouguettaya,~\IEEEmembership{Fellow,~IEEE}
\IEEEcompsocitemizethanks{\IEEEcompsocthanksitem  A. Abusafia, A. Lakhdari, A. Bouguettaya are with the School of Computer Science,
University of Sydney, Australia.
E-mail: \{amani.abusafia, abdallah.lakhdari, athman.bouguettaya\}@sydney.edu.au}}





\IEEEtitleabstractindextext{
\begin{abstract}

We propose a novel spatio-temporal service composition framework for crowdsourcing \textit{multiple} IoT energy services to cater to  \textit{multiple} energy requests. We define a new energy service model to leverage the \textit{wearable-based} energy and wireless power transfer technologies. We reformulate the problem of spatio-temporal service composition to provision multiple energy requests as \textit{a matching problem}. We leverage the \textit{fragmented} nature of energy to offer \textit{partial} services to maximize the utilization of energy services. We propose \textit{EnergyFlowComp}, a modified Maximum Flow matching algorithm that efficiently provisions  IoT energy services to accommodate multiple energy requests. Moreover, we propose \textit{PartialFlowComp}, an extension of the \textit{EnergyFlowComp} approach that considers the \textit{partial-temporal} \textit{overlap}  between services and requests in provisioning. We conduct an extensive set of experiments to assess the effectiveness and efficiency of the proposed framework.

\end{abstract}
\vspace*{-.5em}
\begin{IEEEkeywords} 
EaaS, Energy Services, IoT, Services Composition, Maximum Flow, IoT Services,  Wearable.
\end{IEEEkeywords}}
\maketitle
\IEEEdisplaynontitleabstractindextext
\IEEEpeerreviewmaketitle

\vspace{-10cm}
\IEEEraisesectionheading{\section{Introduction}\label{sec:introduction}}
\vspace{-0.2cm}

\textit{Internet of Things (IoT)} is a paradigm that enables everyday objects (i.e., \emph{things}) to connect to the internet and exchange data \cite{whitmore2015internet}.  IoT devices usually have augmented capabilities, including sensing, networking, and processing \cite{whitmore2015internet}. The number of connected IoT devices is expected to reach 125 billion in 2030 \cite{markit2017internet}. This potential pervasiveness of IoT provides opportunities to \emph{abstract} their capabilities using the \emph{service paradigm} as \emph{IoT services} \cite{lakhdari2020Vision}. IoT services are defined by their functional and non-functional attributes. The functional attributes define the purpose of the service, such as sharing internet access using WiFi. The non-functional attributes are the properties that assess the Quality of Service (QoS), e.g.,  signal strength, reliability, etc. For example, an IoT device owner may offer their WiFi as a hotspot (i.e., service provider) to other nearby IoT devices (i.e., service consumers). A multitude of novel IoT services may be used to enable intelligent systems in several domains, including smart cities, smart homes, and healthcare \cite{abusafia2022services}. Examples of IoT services are WiFi hotspots, environmental sensing, and  energy  services \cite{lakhdari2020composing}. Of particular interest is the use of energy services.\looseness=-1

Energy service, also known as \emph{Energy-as-a-Service (ES)}, refers to the\textit{ wireless power transfer} among nearby IoT devices \cite{lakhdari2020composing}.  We consider a particular set of IoT devices named \textit{wearables}. Wearables refer to anything worn or hand-held like smart shirts, smartwatches, and smartphones \cite{seneviratne2017}. Wearables may harvest energy from natural resources such as kinetic activity, solar power, or body heat \cite{li2023activity}\cite{manjarres2021enhancing}. For instance, a smart shoe using a PowerWalk kinetic energy harvester may produce 10-12 watts on-the-move power\footnote{bionic-power.com}. In this respect,  wearing a PowerWalk harvester may generate energy to charge up to four smartphones from an hour's walk at a comfortable speed. Energy services may be deployed through the newly developed ``Over-the-Air" wireless charging technologies \cite{lakhdari2020composing}\cite{OvertheAirCharger}. Several companies focus on developing the wireless charging technology of IoT devices over a distance, including Xiaomi, Energous, and Cota \cite{lakhdari2021proactive}. For example,  Energous developed a device that can charge up to 3 Watts power within a 5-meter distance to multiple receivers. Another example is Xiaomi’s Mi Air charger which transmits energy wirelessly to nearby IoT devices.\looseness=-1

 \begin{figure*}[!t]
\centering
\includegraphics[width=0.6\linewidth]{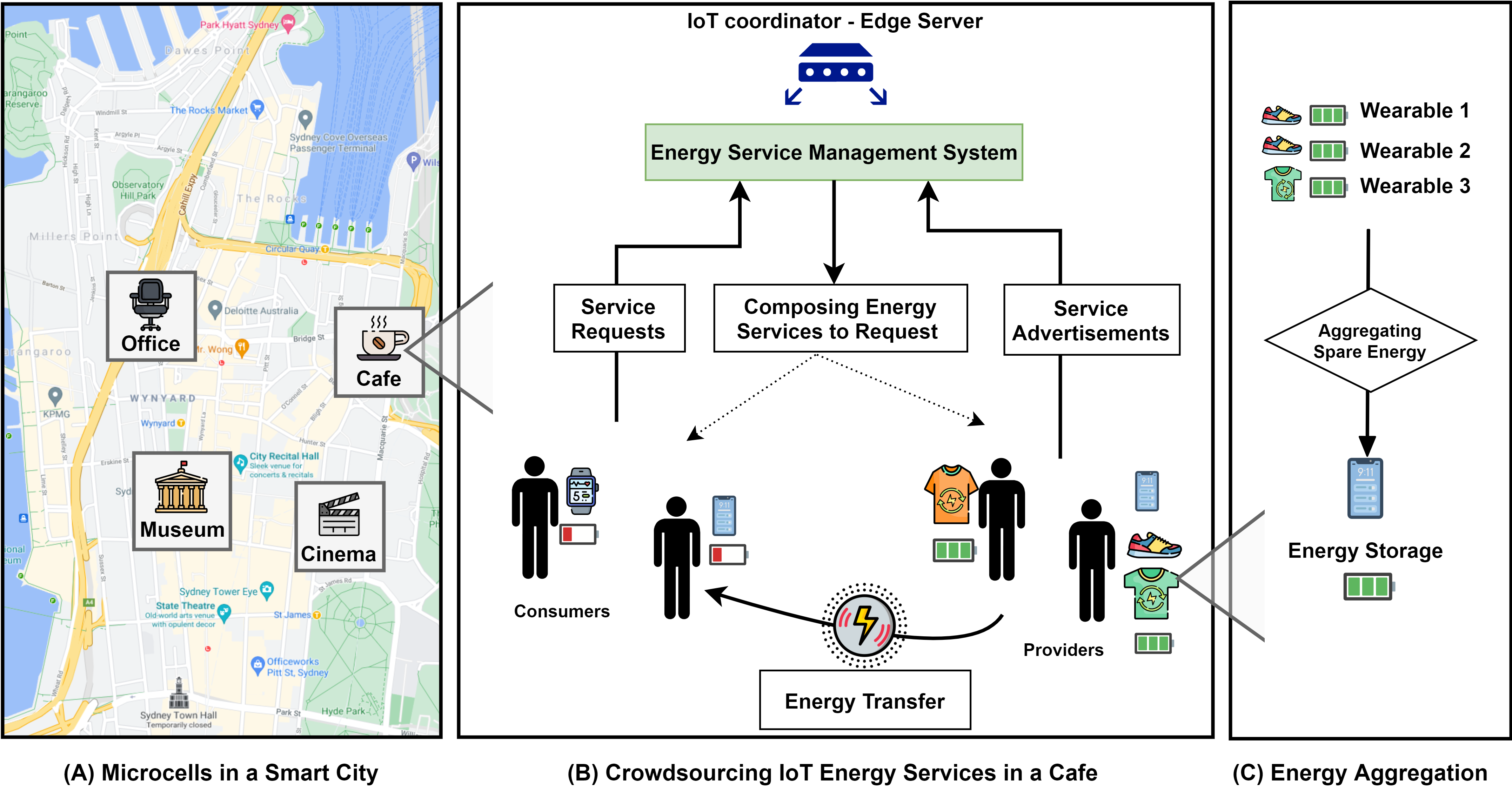}
\setlength{\abovecaptionskip}{1pt}
\setlength{\belowcaptionskip}{-15pt}
\caption{Crowdsourcing IoT energy services scenario}
\label{Scenariofig}
\end{figure*}

\textit{Crowdsourcing} is an efficient way to leverage IoT energy services to create a self-sustained environment \cite{lakhdari2020composing}\cite{abusafia2022services}.  IoT users may \textit{exchange} or \textit{crowdshare} their spare energy to charge nearby IoT devices. Crowdsourced IoT energy services present a \textit{convenient} and  \textit{ubiquitous} power access for IoT users \cite{lakhdari2020composing}\cite{bulut2018crowdcharging}. Indeed,  devices do not need to be tethered to a power point or use power banks. Charging IoT devices wirelessly from a central source usually requires a high-frequency magnetic field to transfer the energy over a distance  \cite{lin2013wireless}. Studies have shown that a strong magnetic field has a harmful impact on humans \cite{baikova2016study}\cite{lin2013wireless}. On the contracts, crowdsourcing IoT energy services enables charging by aggregating energy from multiple close-by devices. As the devices are near, transferring the energy will require a low-frequency magnetic field. Hence, crowdsourcing energy services offer an alternative solution to charge devices wirelessly without compromising users' health.



The proposed crowdsourced IoT energy environment consists of IoT users congregating and moving across \textit{microcells} \cite{lakhdari2020composing}\cite{lakhdari2021proactive}. A microcell is any confined area in a smart city where people may gather (e.g., coffee shops, restaurants, museums, libraries) (see Fig.\ref{Scenariofig}(A)). IoT users are assumed to act as energy providers or consumers (see Fig.\ref{Scenariofig}(B)). The deployment of the energy crowdsourcing environment relies on the willingness of energy providers to participate \cite{lakhdari2020Vision}. We assume that the IoT coordinator offers effective incentives in the form of credits to encourage providers' participation. The credits may be used to receive energy when the providers act as consumers in the future \cite{abusafia2020reliability}\cite{lakhdari2021fairness}. The IoT coordinator is assumed to be deployed one hop away from the energy providers and consumers (e.g., router at the edge) to minimize the communication overhead and latency while advertising energy services and requests.


Energy providers' and consumers' \textit{dynamic} movement  may impact the \textit{balance} between energy services and requests. Energy service providers and consumers may have different preferences regarding location, time, and energy. Therefore, the available energy may be \textit{wasted} by being imprudently allocated to requests. For instance, a  provider may elect to offer a large amount of energy to a small request—the match between a large service and small request results in wasting unused energy. Typically, in crowdsourced environments, the number of service providers is less than the number of consumers \cite{capponi2019survey}. Accordingly, it is crucial to \textit{efficiently} allocate the available scarce energy services \cite{lakhdari2020Vision}. For example, a large service is efficiently utilized by provisioning an existing equal-in-size request. An additional challenge is the likelihood of a single energy service's inadequacy to fulfill the requirement of a consumer due to the limited resources of wearables \cite{lakhdari2020Vision}. In such cases, multiple services may be composed to accommodate a single energy request. Consequently, {\em services composition} is key to efficiently leveraging the available energy in crowdsourced ecosystems.\looseness=-1

Existing energy service composition frameworks mainly focus on  accommodating a \textit{single} request \cite{lakhdari2020Vision}\cite{lakhdari2021fairness} or catering a \textit{single} service \cite{abusafia2020incentive}\cite{abusafia2020reliability}.  These techniques do not consider leveraging \textit{multiple} services to fulfill \textit{multiple} requests simultaneously \cite{lakhdari2021fairness}\cite{lakhdari2020Vision}. This may impede unlocking the full potential of crowdsourcing energy services. This paper focuses on \textit{efficiently matching and composing}  \textit{multiple} energy services to accommodate \textit{multiple} requests simultaneously.




\begin{figure*}[!t]
\centering
\includegraphics[width=0.7\textwidth]{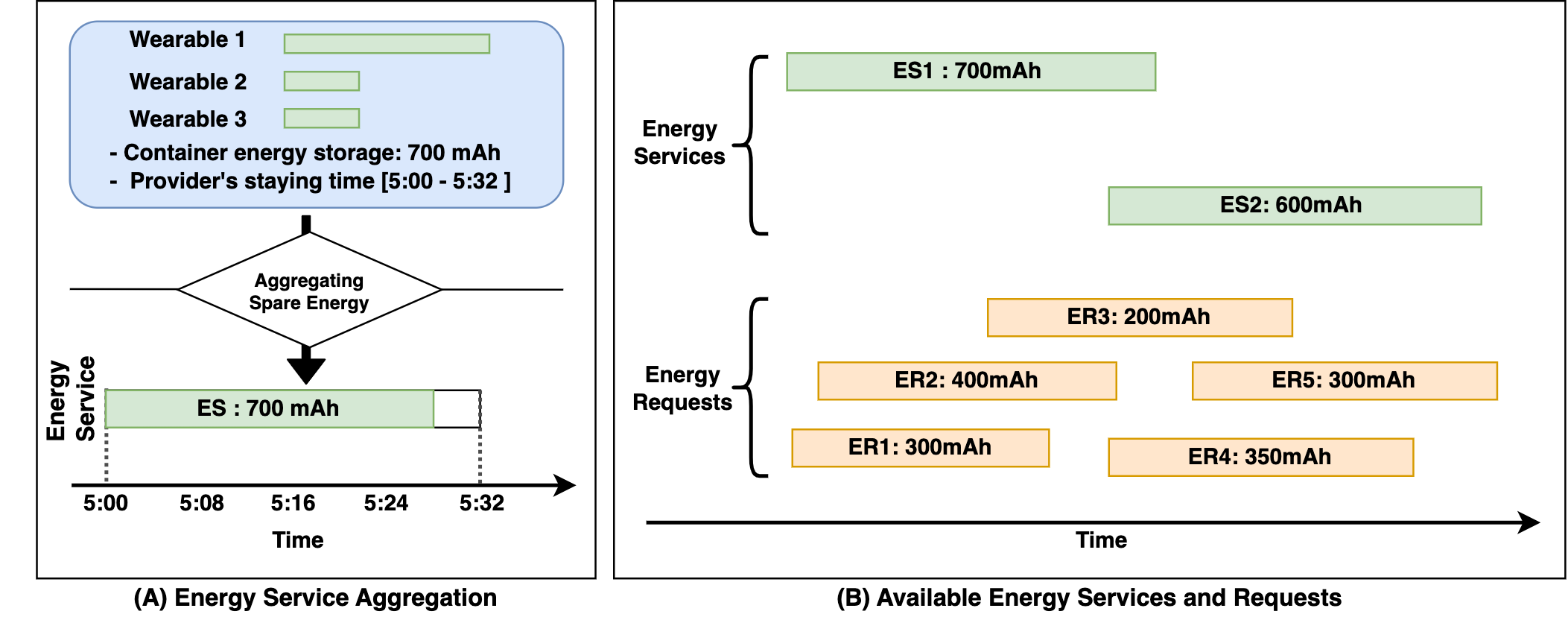}
\setlength{\abovecaptionskip}{1pt}
\setlength{\belowcaptionskip}{-15pt}
\caption{Crowdsourcing wireless energy   example: (A) Phase 1: The proxy aggregates its wearables' spare energy as an energy service (B) Phase 2: The IoT coordinator composes services to requests. }
\label{ExampleOfScenario}
\end{figure*}

We propose a novel spatio-temporal framework to compose \textit{multiple} energy services and fulfill the requirements of \textit{multiple} energy requests within a specified time interval in a confined area. The framework aims to \textit{maximize the utilization of the available energy}. Indeed, in a crowdsourced IoT environment,  it is not guaranteed to fully utilize the available energy services due to the incongruity of the spatio-temporal preferences of energy providers and consumers. Therefore, it is crucial to optimize the available energy services among the existing requests. The framework consists of two levels of composition. The first level allows providers with multiple wearables to aggregate and store their services in a single device (see Fig.\ref{Scenariofig}(C)). This device is used as a proxy to offer energy to other IoT users securely \cite{tarouco2012internet}. The second level is the composition approach to match \textit{multiple} energy services to \textit{multiple} energy requests. We propose \textit{EnergyFlowComp}, a composition approach that utilizes the unique \textit{fragmented} nature of energy to share \textit{partial} service with multiple requests to maximize the utilization of energy \cite{lakhdari2020composing}. Unlike other services, which are atomic (i.e., either fully served or not served at all),  energy services may be divided and shared partially. For instance, a single large energy service may be split to fulfill two smaller requests. Additionally, we propose \textit{PartialFlowComp}, that considers the \textit{partial-temporal} \textit{overlap}  between services and requests in provisioning.\looseness=-1

The main contributions of this paper are:
\begin{itemize}[ noitemsep,nosep,leftmargin=10pt,labelsep=1pt,itemindent=0pt, labelwidth=*]
    \item A novel Aggregated energy service model.
    \item A  two-level composition of \textit{multiple} energy services to fulfill \textit{multiple} requests.
    \item \textit{EnergyFlowComp} a QoS-based \textit{network-flow} approach to match energy services to requests.  
   \item \textit{PartialFlowComp} a {heuristic-based} approach to efficiently allocate energy services to requests.   
   \item A set of extensive experiments on  to show the efficiency and effectiveness of the proposed framework.
\end{itemize}

The paper's remainder is organized as follows. Section \ref{Motivation} presents the motivation scenario. Section \ref{SysMdl} provides an overview of the system model and defines the problem. Section \ref{FrameSec} presents the details of the proposed energy composition framework. Section \ref{ExpSection} evaluates the approach and presents the results. Section \ref{RelatedSection} surveys the related work. Section \ref{ConclusionSection} concludes the paper and highlights future work.\looseness=-1

\vspace{-15pt}
\section{Motivation Scenario}
\label{Motivation}
\vspace{-5pt}


We consider the following motivating scenario to highlight the significance of our work. Let us consider a mall as the general geographical area for crowdsourcing energy services. We divide this area into microcells, where each cell is a confined space (e.g., coffee shop, restaurant, movie theater, etc.) (see  Fig.\ref{Scenariofig}(A)). Each microcell contains IoT devices that provide or receive energy (see Fig.\ref{Scenariofig}(B)). The devices are assumed to be equipped with wireless energy transmitters and receivers such as Energous\footnote{https://www.energous.com/}. All energy requests and advertisements are submitted and processed at the \textit{edge} (e.g., router that is associated with a microcell). The IoT \textit{coordinator} at the edge is responsible for managing and matching the energy services and requests to utilize the available energy efficiently. Additionally, the IoT coordinator is responsible for providing rewards to encourage providers to offer their services  \cite{abusafia2020incentive}.\looseness=-1

We assume \textit{IoT users} are willing participants to provide energy to nearby IoT devices as energy {providers}. Each provider has at least one wearable that harvests energy (e.g., smart shoe) or has spare energy (e.g., smartphone). Energy providers use one of their devices as a \textit{proxy} to \textit{store} the spare energy (see Fig.\ref{Scenariofig}(C)). The proxy is assumed to have computing and energy storage capabilities, i.e., a smartphone. The proxy uses its extra energy storage (i.e., battery) as a container to aggregate and store the spare energy from the provider's wearables. The proxy uses the aggregated energy to define its energy service. We assume the proxy's container is big enough to store all the wearables' spare energy.\looseness=-1 


The process of crowdsourcing energy services consists of two phases: in the first phase, each provider aggregates their wearables' spare energy into a single energy service. For example,  provider 1 would collect and aggregate 700 mAh of energy in its energy storage from its three wearables, e.g., two smart shoes and a smart shirt. (see Fig.\ref{ExampleOfScenario}(A)). In the second phase, the IoT coordinator receives energy services advertisements from the providers and energy requests from the consumers (see Fig.\ref{ExampleOfScenario}(B)). We assume that consumers and providers have different requirements in terms of energy, time availability, and location. The IoT coordinator, then, \emph{coordinates and allocates} the received energy services to requests. \textit{ We focus on optimizing the two phases of the crowdsourcing energy services to maximize the utilization of the available energy}.\looseness=-1

Fig.\ref{ExampleOfScenario} represents a detailed scenario of seven IoT users staying at a coffee shop. Two users are assumed to be willing to provide their energy services, ES1, and ES2, respectively. The other five IoT users request energy from their neighboring IoT devices ER1, ER2, ER3, ER4, and ER5. The advertisement of services and requests includes various information, e.g.,  start time, end time, location of the IoT device, and provided or requested energy amount. Fig.\ref{ExampleOfScenario}(B) illustrates the available services and requests by their timelines. It also shows the amount of available energy. It is challenging to efficiently leverage the available energy and fulfill multiple requests' requirements. Traditional scheduling strategies, such as FCFS and priority-based schedulers, may not be a good fit to provision crowdsourced IoT energy \cite{lakhdari2021fairness}. These strategies fulfill the requirements of each energy request independently and sequentially based on their arrival time. As a result, services may satisfy the requirements of an energy request without being fully utilized, which affects the energy allocation efficiency. For example, assigning ES1 (700 mAh) to ER1 (300 mAh) will use 42.9\% from ES1. Alternatively,  assigning ES1 to ER2 (400 mAh) will utilize  57.1\% from ES1.\looseness=-1

The limited provided energy and time constraints of services and requests represent critical challenges for efficient and maximum provisioning of IoT energy services. We reformulate our service provisioning problem as {\em a  time-constrained optimization problem, i.e., maximizing the allocated energy provided by the available services with respect to their time constraints}. Our proposed approach \textit{EnergyFlowComp} transfers the spatio-temporal services and requests to a Bipartite graph \cite{heineman2016algorithms}. The approach then matches the services to requests using a modified  Maximum Flow algorithm \cite{heineman2016algorithms}. Furthermore, our approach employs the divisibility nature of energy services to maximize energy utilization \cite{lakhdari2020composing}. In addition, we propose \textit{PartialFlowComp}, an extension of  \textit{EnergyFlowComp} that considers the \textit{partial-temporal} \textit{overlap}  between services and requests in provisioning. For example, as ES1 partially overlaps with ER4, it may offer a partial service to part of ER4.\looseness=-1
\vspace{-10pt}
\section{SYSTEM MODEL AND PROBLEM FORMULATION}
\label{SysMdl}

We formally present the definitions of IoT energy services and requests. We then introduce the concept of energy utilization among energy services based on their allocated energy. This work considers a provisioning framework for stationary services and requests to focus only on the temporal constraints in allocating multiple energy services to multiple requests in a microcell within a predefined time interval. The aim is to ensure maximizing the energy provision over a predefined time.\looseness=-1
\vspace{-15pt}
\subsection{Energy Service Model}

We define an energy provider as an IoT user willing to share their devices' spare energy with nearby IoT devices.  We use energy services to abstract the energy-sharing process.  An energy service is the abstraction of the wireless energy delivery from an IoT device to another \cite{lakhdari2018crowdsourcing}.  In the previous energy service model, energy provisioning was from a single energy harvester \cite{lakhdari2020composing}.  We extend the existing energy service model by formally representing the aggregated energy from a set of wearables owned by the IoT user.  The goal is to define the aggregated energy amount, time interval automatically, and spatial location to provide the energy service.  We formally define the  energy service as follows:


\textbf{Definition 1:  Energy service  (ES).} An ES is defined as a tuple of $<EID, PID, W, F, Q>$, where:
\begin{itemize}[ noitemsep,nosep,leftmargin=4pt,labelsep=2pt,itemindent=0pt, labelwidth=*]
    \item $EID$ is a unique identifier for energy service.
    \item $PID$ is a unique identifier for the provider.
    \item $W$ is a set of wearables $\{w_1, ..., w_n\}$ owned by the provider.
    \item $F$ is the function of sharing energy wirelessly.
    \item $Q$ denotes a QoS property of the energy service \cite{lakhdari2018crowdsourcing}.
\end{itemize}

\textbf{Definition 2. Quality attributes of an energy service (QoS).} QoS is a set of parameters that allows energy consumers to distinguish among energy services \cite{lakhdari2020composing}, namely:\looseness=-1
\begin{itemize}[ noitemsep,nosep,leftmargin=4pt,labelsep=2pt,itemindent=4pt, labelwidth=*]
    \item ${ae}$ is the amount of aggregated energy from the provider's wearables that they may share.
    \item ${l}$ is the spatial location of the provider.
    \item ${r}$ is the range for a successful wireless energy transfer.
    \item $[st, et]$ represents the start and end times of the provider's aggregated energy service, respectively.
\end{itemize}

\begin{figure*}[!t]
\centering
\includegraphics[width=0.6\linewidth]{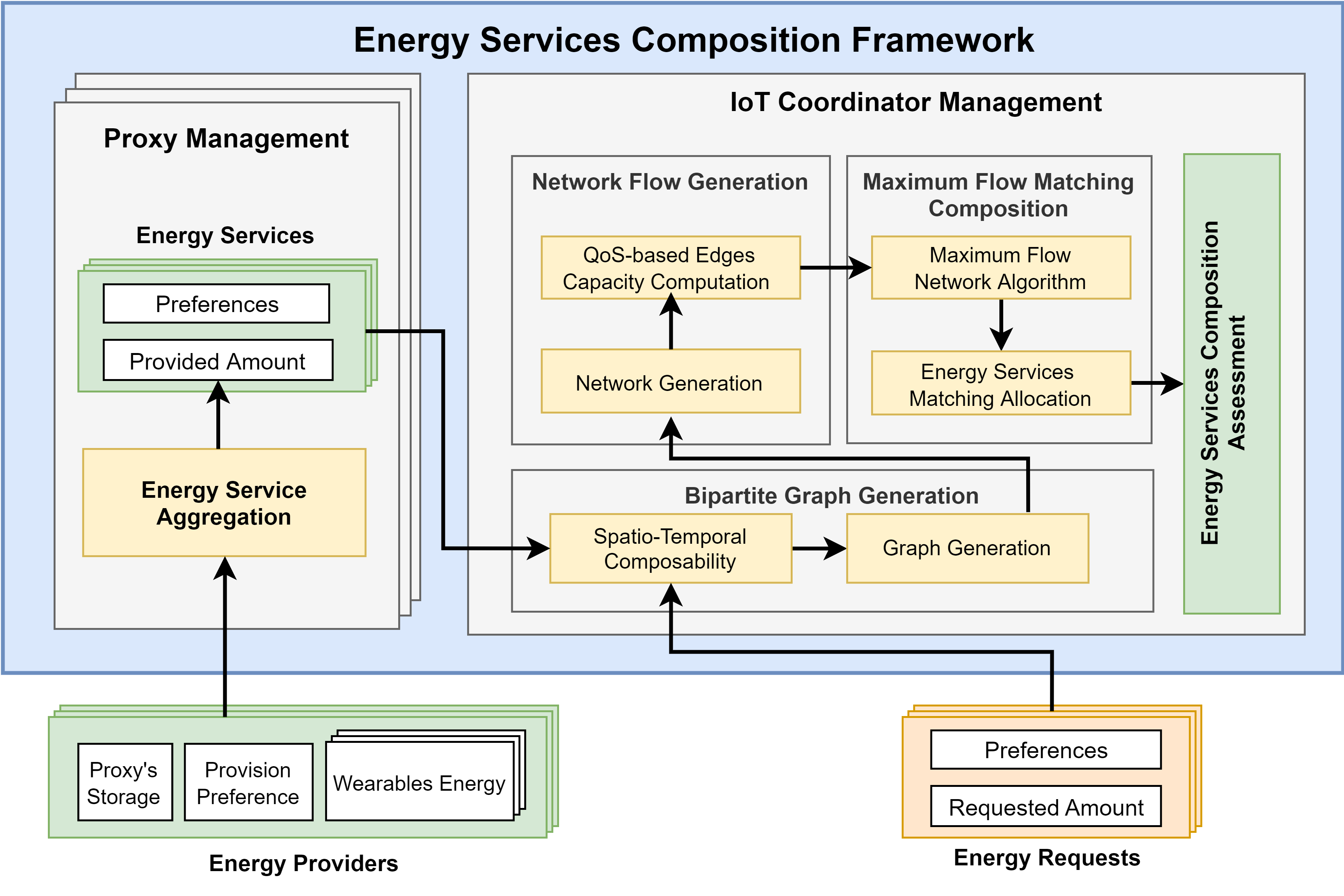}
\setlength{\abovecaptionskip}{-3pt}
\setlength{\belowcaptionskip}{-13pt}
\caption{Wireless energy services framework}
\label{Framework}
\end{figure*}

\vspace{-10pt}
\subsection{Energy Request Model}
\vspace{-3pt}
An energy consumer in the crowdsourced IoT environment is an IoT user who requests energy for their devices from nearby IoT devices. An energy request is defined as the abstraction of the energy consumer requirement according to their spatio-temporal preferences \cite{lakhdari2018crowdsourcing}\cite{abusafia2020incentive}. We formally define an energy service request as follows:

\textbf{Definition 3: Energy Service Request (ER).} An ER is defined as a tuple of $<RID, CID, re, du,l>$, where:
\begin{itemize}[ noitemsep,nosep,leftmargin=4pt,labelsep=2pt,itemindent=0pt, labelwidth=*]
    \item $RID$ is a unique identifier of energy request.
    \item $CID$ is a unique identifier of the consumer.
    \item $re$ is the amount of requested energy by the consumer.
    \item $[st, et]$ is the duration of the consumer's request.
    \item ${l}$ is the spatial location of the consumer.
\end{itemize}
\vspace{-10pt}
\subsection{Energy Utilization and Problem Formulation}
\vspace{-3pt}
The goal of crowdsourcing energy services is to utilize the available energy by efficiently allocating it to the existing energy requests. 
Typically, the available energy services in a crowdsourced IoT environment are limited \cite{lakhdari2021fairness}. Therefore, allocating the available services efficiently is essential to maintain a self-sustained ecosystem \cite{lakhdari2020Vision}. We formally define the energy utilization as follows:\looseness=-1

\textbf{Definition 4: Energy Utilization.} The energy utilization of an $ES$ may be defined as the ratio of the allocated energy to the advertised amount of energy. We measure the Energy Utilization ($EU$) of a composition  using the following:\looseness=-1
\vspace{-10pt}
\begin{equation}
\label{Eq:EU}
\setlength{\abovecaptionskip}{-20pt}
\setlength{\belowcaptionskip}{-20pt}
\begin{split}
     EU(Comp) =  \frac{\sum\limits_{i=1}^n{ES_i.all}}{\sum\limits_{i=1}^n{ES_i.ae}} \; \forall \; ES_i\in \mathcal{ES},\\ \textnormal{\textbf{if }} ES_i.[st,et] \subset T \textnormal{\textbf{ and }} ES_i \textnormal{\textbf{ is serviceable}}
\end{split}
\end{equation}
where $ES_i.all$ is the service's allocated energy to requests,  $ES_i.ae$ is the service's available energy, and $n$ is the number of available energy services $\mathcal{ES}$. Note that a service will be considered if (1) its time duration $[st, et]$ falls within the time window of the composition $T$, and (2) its serviceable, meaning there are energy requests that can utilize that service.\looseness=-1




\textbf{Definition 5: Energy Services Composition problem.}  
We assume that in a microcell, there exists a set of $n$  energy services $\mathcal{ES} = \{ES_1, ES_2, ..., ES_n\}$ and a set of $m$ energy requests $\mathcal{ER} = \{ER_1, ER_2, ..., ER_n\}$ as shown in Fig.\ref{ExampleOfScenario}(B). The $\mathcal{ES} $ will be advertised by providers $P$. Each $ES_i$ is described using the aforementioned Definitions 1 and 2. The energy requests are submitted by consumers $C$. Each energy request $ER_j$ is described using the aforementioned Definition 3. We formulate the problem of allocating $\mathcal{ES}$ to $\mathcal{ER}$ as a time-constrained resource matching problem as follows:\looseness=-1
\begin{itemize}
    \item Maximize $ EU(Comp)$,
\end{itemize}
Subject to :
\begin{itemize}
    \item $ \sum re_j \leq ae_i$  for each $ES_i  \in  \mathcal{ES}$,
    \item $  ER_j.[st,et] \subset ES_i.[st,et]$  for each $ES_i  \in  \mathcal{ES}$.
\end{itemize}
The objective function attempts to optimally allocate the available energy services to the existing requests according to their spatio-temporal features.

\noindent We use the following assumptions to formulate the problem:\looseness=-1

\begin{itemize}[ noitemsep,nosep,leftmargin=1pt,labelsep=2pt,itemindent=0pt, labelwidth=*]
    \item All IoT energy services and requests are deterministic and stationary, i.e., there is prior knowledge about service availability, QoS values, energy requests, time constraints, and demands \cite{lakhdari2018crowdsourcing}.
    \item The IoT coordinator (at the edge) is responsible for {\em batching} the energy requests and energy services from all consumers and providers in a microcell {\em over a predefined period of time}.
    \item There is no energy loss in sharing, i.e., as the technology of wireless power transfer matures, we anticipate that the devices will be able to share more energy, and the energy loss of sharing will become minimal \cite{lakhdari2021fairness}.  
    \item The composition considers the case of multiple providers and multiple consumers.
    \item Wearables have a fixed energy size during the composition.
    \item The proxy's energy container is always sufficient to accommodate the wearables' energy.
    \item The proxy will not use the aggregated energy.
    \item Providers and consumers may have different time windows.
    \item A trust framework has been implemented to preserve the integrity and privacy of the participating IoT devices \cite{abusafia2022Quality}.

\end{itemize}

\vspace{-10pt}

\section{Wireless Energy Sharing Framework}
\label{FrameSec}
In this section, we present the composition framework of crowdsourced energy services. The aim of the composition is to maximize the utilization of the available energy services within a time window by fulfilling nearby requests with respect to their spatio-temporal and energy-related constraints. The framework takes as input the energy services and the energy requests. We assume that energy consumers define their energy requirements and their request duration based on predefined consumption models \cite{lakhdari2020composing}. In our framework, all the available energy requests and services within a predefined time frame are batched. This holistic view allows the IoT coordinator to efficiently allocate the available energy among the existing requests with respect to their time constraints. The framework of the Energy services composition consists of two phases: The  Proxy Management Phase and the IoT Coordinator Management Phase (see Fig.\ref{Framework}). The first phase aggregates the wearables' spare energy of a provider to define their energy service. The second phase matches the available services to requests. 

\begin{figure*}[!t]
\centering
\includegraphics[width=0.8\linewidth]{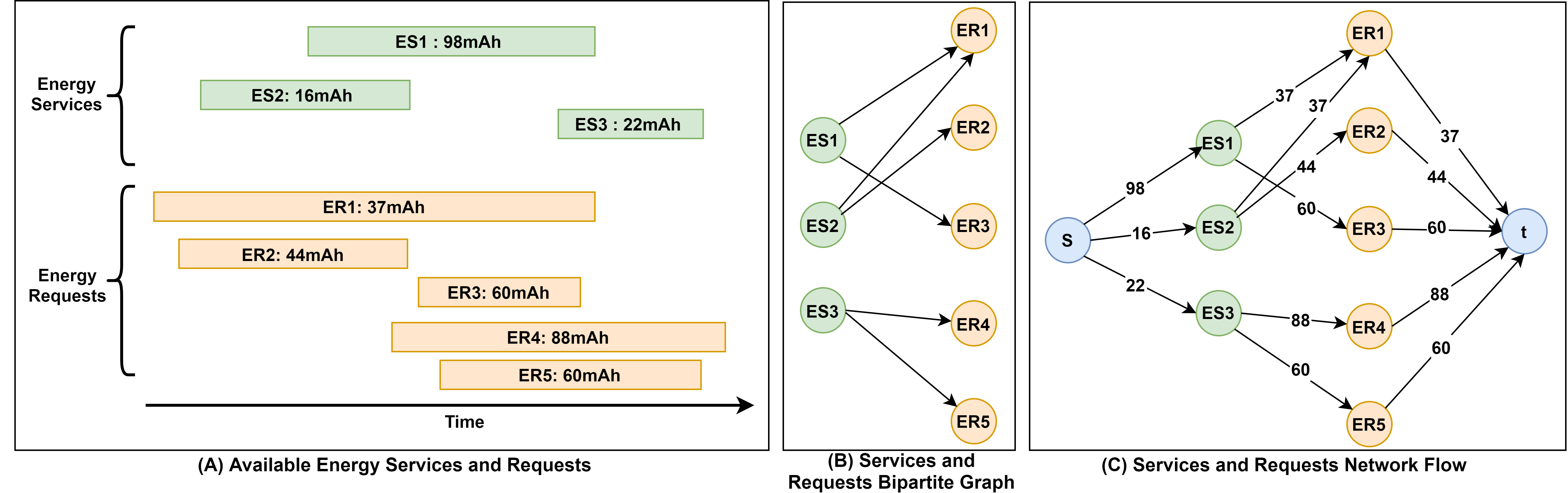}
\setlength{\abovecaptionskip}{-1pt}
\setlength{\belowcaptionskip}{-15pt}
\caption{Bipartite graph generation example}
\label{BG_Example}
\end{figure*}
\vspace{-10pt}
\subsection{\textbf{Phase 1: Proxy Management}} 
\label{Proxy}
The first phase takes as an input the amount of available energy by the provider's wearables, the proxy storage size, and the provider provisioning preferences (e.g., start time and duration of provisioning).  The first phase in the energy sharing process involves aggregating the wearables energy to define an energy service (see Fig.\ref{ExampleOfScenario}(A)).  The aggregation is achieved by transferring the energy from the wearables to the proxy energy storage.  We assume the energy storage is enough to contain all the aggregated energy.  Algorithm \ref{alg:PM} represents the phase in detail.  Algorithm \ref{alg:PM} takes as input the amount of harvested energy by wearables ($\mathcal{E}$) and the temporal preferences of the IoT devices owner ($time$).  The output of the aggregation algorithm is an energy service ($ES$).  Lines [1-2] aggregate the wearables' energy amounts by adding them up.  Lines [3-4] define the spatio-temporal features of the service from the preferences of the owner.  Line 5 defines the energy service ($ES$) using Definitions 1 and 2.\looseness=-1

\vspace{-11pt}
\subsection{Phase 2: IoT Coordinator Management}
The IoT coordinator management phase aims to maximize the utilization of energy by allocating energy services to the corresponding requests. This phase takes as input the advertised aggregated energy services and energy requests. Four steps implement this phase in the following order:  Bipartite Graph Generation, Network Flow Generation, and Maximum Flow Matching composition and the composition assessment. In the following subsections, we discuss these steps in detail, respectively.\looseness=-1

\vspace{-10pt}
\subsubsection{\textbf{Bipartite Graph Generation}}
\label{BGSec}
The Bipartite Graph Generation step aims to present the spatio-temporal composability between the energy services and requests. This step transfers a given set of energy services $\mathcal{ES}$ and requests $\mathcal{ER}$ to a Bipartite Graph ($BG$) (see Fig.\ref{BG_Example}). $BG$ is a directed graph represented by a set of $(U, V, DE)$ where $U$ is a set of nodes that represent the available services, $V$ is a set of nodes that represent the available requests, and $DE$ is a set of directed edges from $U$ to $V$. For example, if two nodes ($u_i$,$v_j$) are connected with an edge $ed_k$, it indicates that the service  $ES_i$ can fulfill the request $ER_j$. For instance, in Fig.\ref{BG_Example}(A), the energy services $ES_2$ falls within the duration of the energy request $ER1$. Therefore, an edge will be added in the graph $BG$ between the node of $ES_2$ and $ER_1$ (see Fig.\ref{BG_Example}(B)).
\begin{algorithm}[t!]
    \footnotesize
     \renewcommand{\algorithmicrequire}{\textbf{Input:}}
    \renewcommand{\algorithmicensure}{\textbf{Output:}}
    \caption{Energy Services Aggregation}
    \label{alg:PM}
    \begin{algorithmic}[1]
        \Require
        $\mathcal{E}, \;time$
        \Ensure $ES$
        \Statex \textbf{//Phase 1: Proxy Management}
        \For{$e_i\; \textbf{\textit{in}}\; \mathcal{E}$} 
        \State energy\_amount $\gets$ energy\_amount + aggregate ($e_i$)
        \EndFor
        \State {time\_preference $\gets$ Define\_Time\_Interval(\textit{time})}
        \State {location $\gets$ Detect\_Location()}
        \State $ES\gets$Define\_Energy\_Service ( 
        energy\_amount, time$ $\_preference, location) 
        \State \Return $ES$ \label{endph2} \label{startph2}
    \end{algorithmic}
\end{algorithm}
\setlength{\textfloatsep}{4pt}
\\
\begin{algorithm}[!t]
    \footnotesize

     \renewcommand{\algorithmicrequire}{\textbf{Input:}}
    \renewcommand{\algorithmicensure}{\textbf{Output:}}
    \caption{ Services Bipartite Graph Generation }
    \label{alg:BGG}
    \begin{algorithmic}[1]
        \Require
        $\mathcal{ES},\; \mathcal{ER}, T$
        \Ensure $BG$
        
        \State $\mathcal{ES} =$ Temporal\_filter ($\mathcal{ES}, T$)
        \State $\mathcal{ER} =$ Temporal\_filter ($\mathcal{ER}, T$)
        \State $BG = \{\}$
        \For{$ES_i\; \textbf{\textit{in}}\; \mathcal{ES}$}
            \State $BG.$Add\_node($ES_i.EID, ES_i, type:U$)
        \EndFor 
        \For{$ER_j\; \textbf{\textit{in}}\; \mathcal{ER}$}
            \State $BG.$Add\_node($ER_j.RID, ER_j, type:V$)
        \EndFor
         \Statex \textbf{// Spatio-Temporal Composability} 
            \For{$ES_i\; \textbf{\textit{in}}\; \mathcal{ES}$}
                \For{$ER_j\; \textbf{\textit{in}}\; \mathcal{ER}$}
                    \If{Nearby($ES_i.l, ER_j.l$)}
                         \If{$ES_i.[st, et] \subseteq ER_j.[st, et] \parallel ER_j.[st, et] \subseteq ES_i.[st, et]  $}
                            \Statex \;\;\;\;\; \; \; \; \;\;\;\;\; \; \; 
\textbf{// Graph Generation} 
                            \State $w = Er_j.r$
                            \State  $BG.$Add\_edge($ES_i.EID,ER_j.RID,w$)

                        \EndIf
                    \EndIf
                \EndFor
            \EndFor
        \Return $BG$
    \end{algorithmic}
\end{algorithm}
Algorithm \ref{alg:BGG} represents the step of bipartite graph generation in detail. Algorithm \ref{alg:BGG} takes as an input the available energy services $\mathcal{ES}$ and requests $\mathcal{ER}$ and the time window of the composition $T$. The output of the algorithm is a Bipartite Graph ($BG$). Lines 1-2 select the services and requests that fall within the time window $T$. Lines 3-7 add the available services and requests as nodes to the graph. Then for each energy service and request, the algorithm checks if the service and request are spatially composable (Lines 8-11). A service and a request are considered spatially composable, if and only if, the distance between them is less than the required distance for a successful energy wireless transfer \cite{lakhdari2020composing}.   Additionally, the algorithm checks the temporal composability between each service and a request (Line 11). We extend the definition of temporal composability by \cite{lakhdari2020composing} to consider a service and a request are temporally composable if the duration of the service falls within the duration of the request or vice versa \cite{lakhdari2020composing}. If a service and a request are spatio-temporally composable, then an edge will be added to the graph (Lines 12-13). The edge connects the node of the service to the node of the request indicating that the service may fulfill the energy request. Line 12 computes the edge's weight $w$, which is equal to the required amount of energy by the request. Line 13 adds the edge between the service's node and the request's node in the bipartite graph $BG$. The added edge indicates that the service node may fulfill the energy request up to its $w$. The output of the algorithm is a bipartite graph. The graph represents the spatio-temporal composability between the energy services and requests.

\vspace{-8pt}
\subsubsection{\textbf{Network Flow Generation}}
\label{NFSec}
Traditionally, network flow has been used to solve bipartite graph matching \cite{kleinberg2006algorithm}. In this step, we aim to generate a network flow from the bipartite graph resulting from Algorithm \ref{alg:BGG} (see Fig.\ref{BG_Example}(C)). The network flow is generated with respect to the quality of services (QoS) of both services and requests. The network flow generation step consists of two main components: (1) Network Generation and (2) QoS- based edges' capacity computation. The step of network generation converts the bipartite graph to a network by connecting all the nodes to a source and a sink. The step of capacity computation uses the QoS attributes to calculate the capacity of each edge.\looseness=-1
\begin{figure*}[!t]
\centering
\includegraphics[width=0.8\linewidth]{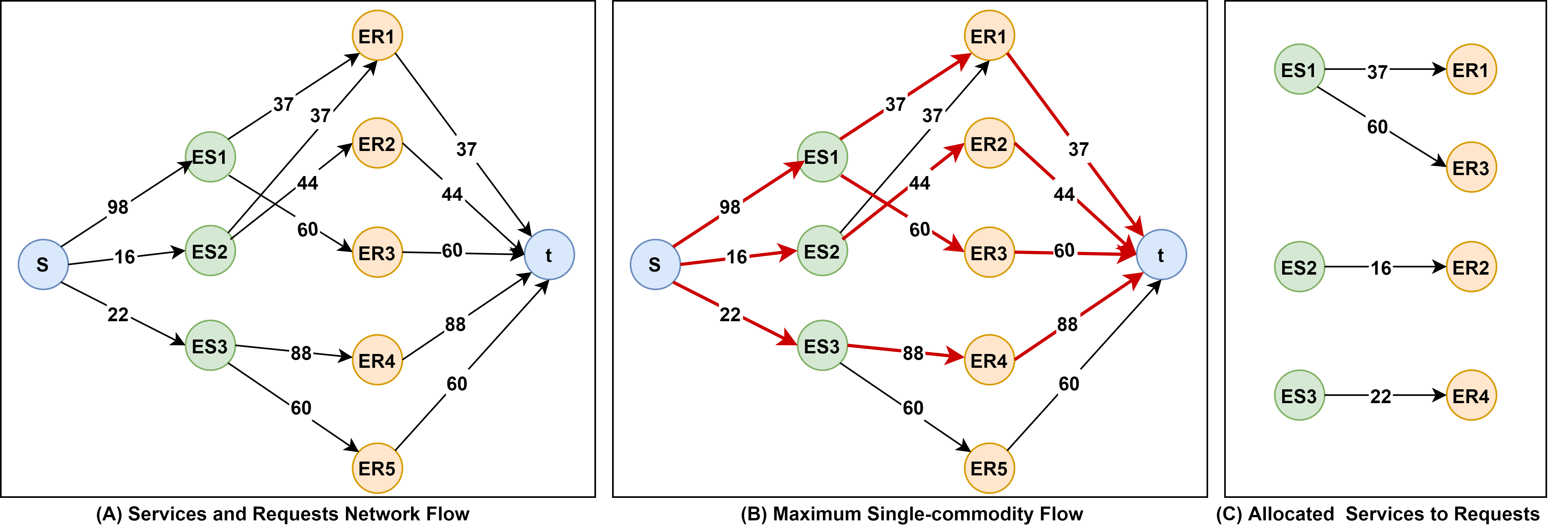}
\setlength{\abovecaptionskip}{1pt}
\setlength{\belowcaptionskip}{-15pt}
\caption{Maximum flow matching composition}
\label{MF_Example}
\end{figure*}

Algorithm \ref{alg:NFG} represents the services network flow generation in detail. Algorithm \ref{alg:NFG} takes as an input the generated bipartite graph $BG$, available energy services $\mathcal{ES}$ and requests $\mathcal{ER}$. The output of the algorithm is a network flow $NF$. Lines 1-3 initialize the network and add a source $s$  and a sink $t$. Lines 4-13 traverse every node in the bipartite graph $BG$ to add it to the network (Line 5). Additionally, if the node's type is \textit{`U'} (i.e., the node represents a service), then the node will be connected to the source $s$ (Lines 6-9). In addition, the capacity of the edge will be the amount of energy that may be provided by the service (Line 8). Moreover, if the node's type is \textit{`V'} (i.e., the node represents a request), then the node will be connected to the sink $t$ (Lines 10-13). In addition, the edge's capacity will be the amount of energy that may be inquired by the request (Line 12). Lastly, the algorithm retrieves the neighbour nodes of every \textit{`U'} node in $BG$ to add an edge between them in the network flow $NF$. Recall that the \textit{`U'} node is a service whose neighbors are requests (Lines 14-16). Line 18 computes the edge's capacity $capacity$, which is equal to the edge's weight $w$ in the bipartite graph $BG$. Recall that  $w$  is equal to the required amount of energy by the request. Line 19 adds the edge between the node and its neighbour to the network flow $NF$. The added edge indicates that the service node may fulfill the energy request up to its $capacity$. The algorithm's output is a network flow representing the energy flow from services to requests with respect to their QoS attributes.
 \vspace{-8pt}
\subsubsection{\textbf{Maximum Flow Matching Composition}}
\label{MaxFlowSec}

 In this step, we propose a Maximum-Flow matching Composition (\textit{EnergyFlowComp}). Our approach allocates energy services to requests using a modified maximum flow algorithm. The Maximum-Flow matching Composition \textit{EnergyFlowComp} consists of two components: (1) Maximum-Flow network algorithm and (2) Energy services matching allocation (see Fig.\ref{Framework}). The first component of the composition uses a maximum flow algorithm to find a maximum single-commodity flow. For instance, using the previous example of Fig.\ref{BG_Example}, the result of the first component in the Maximum-Flow matching Composition is shown in red in Fig.\ref{MF_Example}(B). The arrows indicate the path to get the maximum flow in the network. In our context, the maximum flow will be the maximum energy services that can be allocated to the existing requests. Note that a selected edge between two nodes in the network flow does not mean the full use of its capacity. For instance, in Fig.\ref{MF_Example}(B), the capacity of the edge between ES2 and ER2 is 44. However, the capacity of the edge between the source $s$ and ES1 is 16, which means the available energy by ES1 is 16. Hence, the allocated energy is 16, as shown in the edge between ES2 and ER2 in Fig.\ref{MF_Example}(C). The second component of the composition traverses the resulted flow to assign services to requests (see Fig.\ref{MF_Example}(C)).
 
  \begin{algorithm}[!t]
     \footnotesize
     \renewcommand{\algorithmicrequire}{\textbf{Input:}}
    \renewcommand{\algorithmicensure}{\textbf{Output:}}
    \caption{Services Network Flow Generation }
    \label{alg:NFG}
    \begin{algorithmic}[1]
        \Require 
        $BG, \;\mathcal{ES},\; \mathcal{ER}$
        \Ensure $NF$
        \State $NF = \{\}$
        \Statex // Add the source node of the network
        \State $NF.$Add\_node($s$)
         \Statex // Add the sink node of the network
        \State $NF.$Add\_node($t$)
        \For{$node_i\; \textbf{\textit{in}}\; BG$}
            \State $NF.$Add\_node($node_i$)
           \If{$node_i.type == `U'$}
             \State $ es = Get\_service(\mathcal{ES},node_i.EID)$
             \State $capacity = es.ae$
             \State $NF.$Add\_edge($s,node_i,capacity$)
            \EndIf
            \If{$node_i.type == `V'$}
             \State $ er = Get\_request(\mathcal{ER},node_i.RID)$
             \State $capacity = er.re$
             \State $NF.$Add\_edge($node_i,t,capacity$)
           \EndIf
        \EndFor 
        
        \For{$node_i\; \textbf{\textit{in}}\; BG$}
           \If{$node_i.type == `U'$}
             \State $ neighbours = Get\_connected\_nodes(node_i)$
             \For{$n_j\; \textbf{\textit{in}}\; neighbours$}
             
             \State $capacity = n_j.w$
             \State $NF.$Add\_edge($node_i,n_j,capacity$)
             \EndFor
            \EndIf
        \EndFor 
        \Return $NF$
    \end{algorithmic}
\end{algorithm}
\setlength{\textfloatsep}{4pt}

The \textit{EnergyFlowComp} approach is described in details in Algorithm \ref{alg:ESC}.  Algorithm \ref{alg:ESC} takes as an input the providers' provisioning preferences $\mathcal{P}$, available energy requests $\mathcal{ER}$, and the composition's time window $T$. The output of the algorithm is a composition of services allocated to the requests $Comp$. Lines 1-4 define the available energy services $\mathcal{ES}$ based on the available providers $\mathcal{P}$. For each provider $P_i$, our approach computes its services using the energy services aggregation algorithm described in Section \ref{Proxy}. Line 5 generates the bipartite graph $BG$ from the available energy services $\mathcal{ES}$, requests $\mathcal{ER}$, and the time window of the composition $T$. The bipartite graph $BG$ will be used to match services to requests. The bipartite graph generation is presented in Section \ref{BGSec}. Line 6 generates the network flow  $NF$ from the generated bipartite graph $BG$, available energy services $\mathcal{ES}$ and requests $\mathcal{ER}$. The network flow  $NF$ will be used to run a maximum flow algorithm to get the best match of services to requests. The network flow generation is presented in Section \ref{NFSec}. Line 7 uses a maximum flow algorithm to find a maximum single-commodity flow. There are several state-of-the-art algorithms to compute the maximum flow of a network, including- Preflow–Push, Edmonds–Karp, and Boykov-Kolmogorov \cite{kleinberg2006algorithm}. All these algorithms result in the same output; however, they differ in time complexity. Therefore, we used the Preflow–Push algorithm, as it is considered one of the most efficient maximum flow algorithms \cite{kleinberg2006algorithm}. Line 8 traverses the resulted flow to assign services to requests. The algorithm's output is a composition that matches services to requests with respect to their QoS attributes.\looseness=-1

\vspace{-8pt}
\subsubsection{\textbf{Energy Services Composition Assessment}}
The last step in the framework is the energy services composition assessment. In what follows, we define the metrics to assess the efficiency of our proposed framework.
\vspace{-8pt}
\subsubsection*{\textbf{Energy Utilization}}
Recall that our composition's objective is to maximize available energy utilization. Typically, the available energy services in a crowdsourced IoT environment are limited \cite{lakhdari2021fairness}. efficiently allocating the available services is essential to maintaining a self-sustained ecosystem \cite{lakhdari2020Vision}. We measure the Energy Utilization ($EU$) using Eq.\ref{Eq:EU}.


\vspace{-8pt}
\subsubsection*{\textbf{Fulfillment Rate}}
We define the \textit{Fulfillment Rate} ($FR$) as the percentage of the required energy that has been fulfilled. Intuitively, better energy allocation and utilization may result in a better fulfillment rate. Therefore, a higher fulfillment rate indicates efficiency in fulfilling the demanded energy. The \textit{Fulfillment Rate} $FR$ of a composition may be computed as follows:
\vspace{-8pt}
\begin{equation}
\label{Eq:FR}
\begin{split}
     FR(Comp) =  \frac{\sum\limits_{i=1}^n{ES_i.al}}{\sum\limits_{j=1}^m{ER_j.re}} \; \forall \; ER_j\in \mathcal{ER},\\ \textnormal{\textbf{if }} ES_i.[st,et] \subset T \textnormal{\textbf{ and }} ES_i \textnormal{\textbf{ is serviceable,}}\\
     \textnormal{\textbf{and if }} ER_j.[st,et] \subset T
     \textnormal{\textbf{ and }} ER_j \textnormal{\textbf{ is serviceable}}
\end{split}
\end{equation}

\begin{figure*}[!t]
\centering
\includegraphics[width=0.6\linewidth]{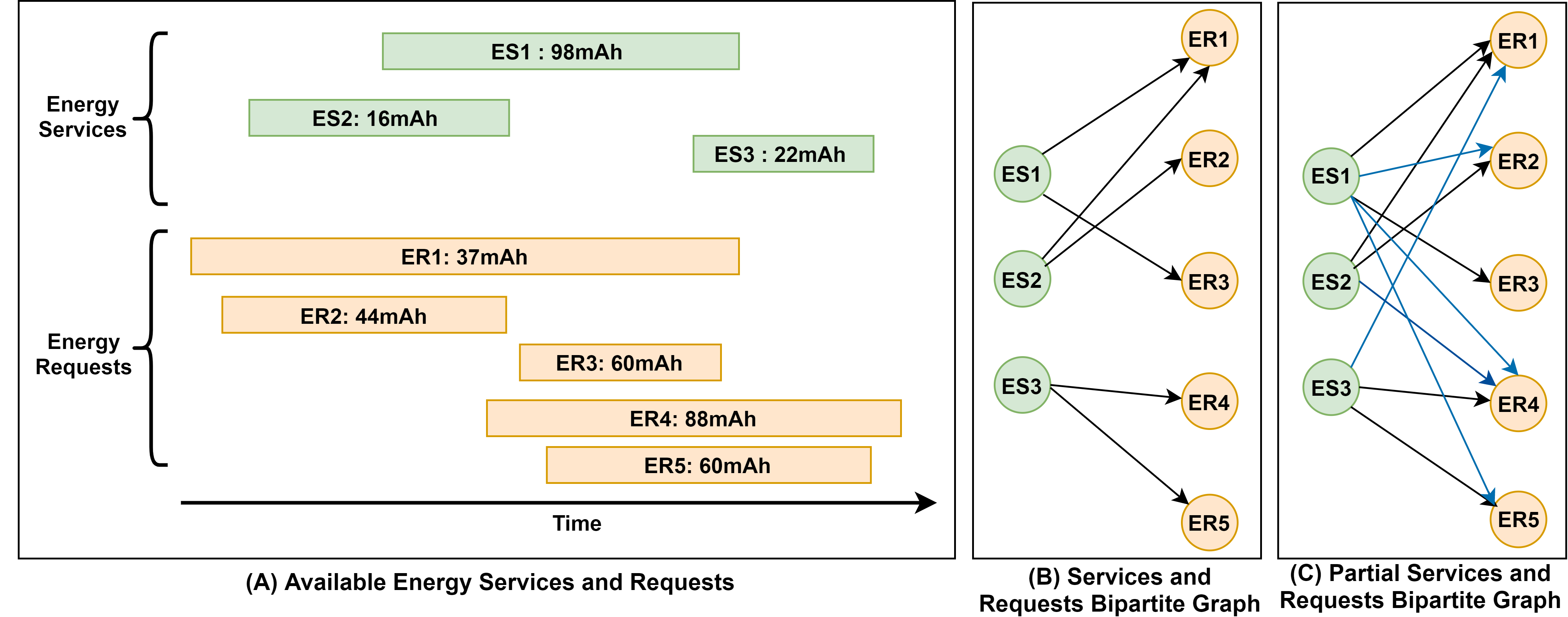}
\setlength{\abovecaptionskip}{-1pt}
\setlength{\belowcaptionskip}{-10pt}
\caption{Example of bipartite graph and partial bipartite graph }
\label{PG_Example}
\end{figure*}
\begin{figure*}[!t]
\centering
\includegraphics[width=0.6\linewidth]{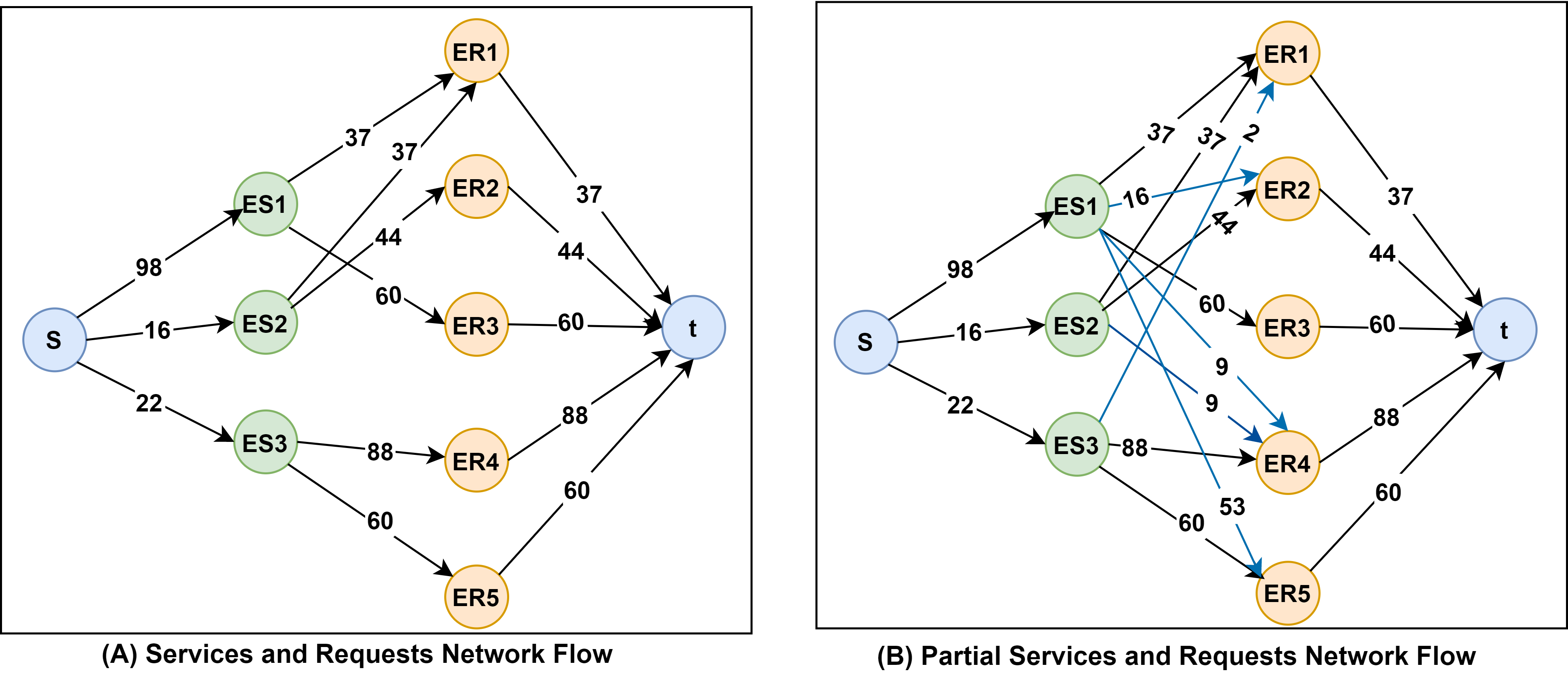}
\setlength{\abovecaptionskip}{-1pt}
\setlength{\belowcaptionskip}{-15pt}
\caption{Example of network flow and partial network flow }
\label{PNF_Example}
\end{figure*}
where $ES_i.all$ is the service's allocated energy to requests, $n$ is the number of available energy services $\mathcal{ES}$, $ER_j.re$ is the required energy by request, and $m$ is the number of available energy requests $\mathcal{ER}$. Note that a service and, similarly, a request will be considered if (1) its time duration $[st, et]$ falls within the time window of the composition $T$, and (2) its serviceable, meaning there are energy requests or services that can utilize or fulfill that service.\looseness=-1
 \begin{algorithm}[!t]
 \footnotesize
     \renewcommand{\algorithmicrequire}{\textbf{Input:}}
    \renewcommand{\algorithmicensure}{\textbf{Output:}}
    \caption{ Energy Services Matching Composition }
    \label{alg:ESC}
    \begin{algorithmic}[1]
        \Require
        $\mathcal{P},\; \mathcal{ER}, T$
        \Ensure $Comp$
        \Statex \textbf{// Energy Services Aggregation }  
       \State $\mathcal{ES}= \{\}$ 
        \For{$P_i\; \textbf{\textit{in}}\; \mathcal{P}$}
            \State $ES$ = Energy\_ Services\_Aggregation($P_i$)
            \State $\mathcal{ES}.append(ES)$
        \EndFor
        \Statex  \textbf{// Bipartite Graph Generation}     
        \State $BG$ = Services\_Bipartite\_Graph\_Generation ($\mathcal{ES},\; \mathcal{ER}, T$)
        \Statex  \textbf{// Network Flow Generation}     
        \State $NF$ = Services\_Network\_Flow\_Generation ($BG, \mathcal{ES},\; \mathcal{ER}$)
        \Statex  \textbf{// Energy Services Matching Allocation}     
        \State $flow $ = Maximum\_Flow\_Network($NF$)
        \State $Comp$ = Energy\_Services\_Allocation($flow,\mathcal{ES},\mathcal{ER}$)

        \Return $Comp$
    \end{algorithmic}
\end{algorithm}
\setlength{\textfloatsep}{4pt}

\vspace{-8pt}
\subsubsection*{\textbf{Consumers and Providers' Satisfaction}} 

Maximizing the energy allocation of the available energy service may {\em satisfy} more consumers by fulfilling their requests. In addition, it may {\em satisfy} more providers by using their services and thereby getting more rewards \cite{abusafia2020incentive}. Typically in resource allocation, consumers may not be satisfied with an efficient allocation if the available resources are limited \cite{radunovic2007unified}. However,  the efficient allocation of energy services in a crowdsourced IoT environment is claimed to increase consumers' satisfaction \cite{lakhdari2021fairness}. We assess the efficiency of our framework by measuring the satisfaction of consumers. The satisfaction  of an energy consumer $j$ is gauged  by  their perception of the allocated energy $ES_i.all$  to their request $ER_j.re$. Therefore, the used equation for computing the consumer satisfaction $CS_i$ is presented as follows:
\vspace{-8pt}
\begin{equation}\label{Eq:CS}
CS_i =
\begin{cases}
    \frac{ES_i.all}{ER_j.re\times Sf}
    ,& \text{if } ES_i.all\leq ER_j.re\\
    1,              & \text{otherwise}
\end{cases}\vspace{-5pt}
\end{equation}

where $ES_i.all$ is the service's allocated energy to the request, $ER_j.re$ is the required energy by a request, and $Sf$ is the satisfaction threshold defined by the total available services \cite{lakhdari2021fairness}. Due to the limited resources in the crowdsourced IoT energy environment, we consider consumers satisfied if they received $Sf\%$ of what they requested \cite{lakhdari2021fairness}.\looseness=-1 

Similarly, we gauge the providers' satisfaction by their perception of the utilized energy $ES_i.all$  from their available services  $ES_i.ae$. The higher the utilization of their services, the higher their rewards \cite{abusafia2020incentive}. Hence, we also assess our framework's efficiency by measuring the providers' satisfaction. The satisfaction  of an energy provider $PS_i$ is computed as follows:\looseness=-1
\vspace{-5pt}
\begin{equation}\label{Eq:PS}
PS_i =
\begin{cases}
    \frac{ES_i.all}{ES_i.ae\times Sf}
    ,& \text{if } ES_i.all\leq ES_i.ae\\
    1,              & \text{otherwise}
\end{cases}\vspace{-5pt}
\end{equation}

where $ES_i.all$ is the service's allocated energy to fulfill requests, $ES_i.ae$ is the size of the energy service, and $Sf$ is the satisfaction threshold defined by the total available services \cite{lakhdari2021fairness}. Due to the temporal constraints of services and requests resources in the crowdsourced IoT energy environment, we consider providers satisfied if they offered $Sf\%$ of their services \cite{lakhdari2021fairness}.\looseness=-1 

\vspace{-6pt}
\subsubsection*{\textbf{Global Consumers and Providers' Satisfaction}}
\label{entropyS_Sec}
We adopt the metric proposed by \cite{lakhdari2021fairness} to define a global satisfaction metric for all providers and consumers. The global consumers' satisfaction is computed by estimating the overall proximity score among the satisfaction scores of all consumers \cite{basik2018fair}\cite{lakhdari2021fairness}. An efficient services composition is reflected by the \textit{closeness} of the satisfaction scores among consumers  (i.e.,  consumers have very similar satisfaction scores). In contrast, the {\em sparsity}  of consumers' satisfaction scores indicate an inefficient service composition. Therefore, similar to \cite{lakhdari2021fairness}, the global satisfaction of consumers $\mathcal{CS}$ may be captured using the {\em information entropy} \cite{shannon1948mathematical}. The information entropy  measures the disorder degree of all requests $ER_j \in \mathcal{ER}$  based on their satisfaction score $CS_j$ as follows:\looseness=-1   
\vspace{-8pt}
\begin{equation} \label{Eq:CSE}
 H(\mathcal{CS}) = -\sum\limits_{j=1}^m CS_j \log_2 CS_j
 \vspace{-8pt}
 \end{equation}
 
 where $m$ is the number of available energy requests $\mathcal{ER}$ and $CS_j$ is a consumer $j$'s satisfaction computed using Eq.\ref{Eq:CS}.
 
Similarly, the global satisfaction of providers $\mathcal{PS}$ may be captured using the {\em information entropy} \cite{shannon1948mathematical}. The information entropy  measures the disorder degree of all services $ES_i \in \mathcal{ES}$  based on the providers' satisfaction score $PS_i$ as follows:   
\vspace{-8pt}
\begin{equation} \label{Eq:PSE}
 H(\mathcal{PS}) = -\sum\limits_{i=1}^n PS_i \log_2 PS_i
 \vspace{-8pt}\end{equation}
 
 where $n$ is the number of available energy services $\mathcal{ES}$ and $PS_i$ is a provider's $i$'s satisfaction computed using Eq.\ref{Eq:PS}.

\begin{table*}
\setlength{\abovecaptionskip}{1pt}
\setlength{\belowcaptionskip}{-8pt}
\centering
\small
\caption{Parameters of the experiments settings}
\begin{tabular}{|l|c|c||l|c|c|}
\hline
\multicolumn{3}{|c||}{\footnotesize{Energy services}}                                                                                                          & \multicolumn{3}{c|}{\footnotesize{Energy Requests}}                                                                                                                                                                                  \\ \hline
\multicolumn{1}{|c|}{\footnotesize{QoS}} & \footnotesize{Dataset}                                                              & \footnotesize{Value}                                                       & \multicolumn{1}{c|}{\begin{tabular}[c]{@{}c@{}}\footnotesize{Query parameters}\end{tabular}} & \footnotesize{Dataset}                                                              & \footnotesize{Value}                                                       \\ \hline
\footnotesize{Start time}                & \footnotesize{IBM Cafe}                                                                 & \footnotesize{Check-in}                                                    & \footnotesize{Start time}                                                                        & \footnotesize{IBM Cafe}                                                                 & \footnotesize{Check-in}                                                    \\ \hline
\footnotesize{End time}                  & \footnotesize{Wireless Energy Dataset}                                                               & \footnotesize{Transfer Duration}                                                      & \footnotesize{End time}                                                                     & \begin{tabular}[c]{@{}c@{}}\footnotesize{Wireless Energy Dataset}\end{tabular}              & \begin{tabular}[c]{@{}c@{}}\footnotesize{Transfer Duration}\end{tabular}     \\ \hline
\footnotesize{Energy capacity}           & \begin{tabular}[c]{@{}c@{}}\footnotesize{Wireless Energy Dataset}\end{tabular} & \begin{tabular}[c]{@{}c@{}}\footnotesize{Provided Energy Amount}\end{tabular} & \footnotesize{Energy capacity}                                                                   & \begin{tabular}[c]{@{}c@{}}\footnotesize{Wireless Energy Dataset}\end{tabular} & \begin{tabular}[c]{@{}c@{}}\footnotesize{Requested Energy Amount}\end{tabular} \\ 
\hline
\end{tabular}
\label{tab:simparam}
\end{table*}

\vspace{-10pt}
\subsubsection{\textbf{Partial Bipartite Graph Generation}}
\label{PBGSec}
In this step, we propose a\textit{ Partial-based}  Bipartite Graph Generation (\textit{PartialGraph}). The \textit{PartialGraph} extends our proposed bipartite graph generation approach to include the \textit{partially overlapping} services and requests.  Considering the partial overlap between services and requests increases the utilization of the available energy.  For example, in Fig.\ref{PG_Example}(A), $ES1$ and $ER2$ are partially overlapping. Using the  bipartite graph generation approach described in Algorithm \ref{alg:BGG} will not connect $ES1$ and $ER2$ (see Fig.\ref{PG_Example}(B)). However, using the partial  bipartite graph generation approach described in Algorithm \ref{alg:PBGG}, $ES1$ and $ER2$ will be connected (see Fig.\ref{PG_Example}(C)). The blue arrows in Fig.\ref{PG_Example}(C) represent all additional edges between services and requests.  The additional edges will result in a more connected network flow (see Fig.\ref{PNF_Example}).\looseness=-1

Algorithm \ref{alg:PBGG} represents the \textit{PartialGraph} in detail. Algorithm \ref{alg:PBGG} takes as an input the available energy services $\mathcal{ES}$ and requests $\mathcal{ER}$ and the time window of the composition $T$. The algorithm's output is a partial bipartite graph ($PBG$). Line 1 generates the bipartite graph $PBG$ from the available energy services $\mathcal{ES}$, requests $\mathcal{ER}$, and the time window of the composition $T$. The bipartite graph generation is presented in Section \ref{BGSec}. Lines 2-8 extend the bipartite graph generation to include partial services. For each energy service and request, the algorithm checks if the service and request are spatially composable (Lines 2-4). A service and a request are considered spatially composable, if and only if, the distance between them is less than the required distance for a successful energy wireless transfer  \cite{lakhdari2020composing}. Additionally, the algorithm checks the \textit{partial temporal overlap} between each service and a request (Line 5).  If a service and a request are partially overlapping in time, then an edge will be added to the graph (Lines 5-8).  Similar to the bipartite graph approach, the edge connects the service's node to the request's node, indicating that the service may fulfill  the energy request. Line 6 computes the duration of overlap between the service and the request. Line 7 computes the edge's weight, which is equal to the part of the request that overlaps with the service, i.e., the required amount of energy by the request within the overlap duration. Line 8 adds the edge between the service node and the requesting node in the bipartite graph $PBG$. The added edge indicates that the service node may fulfill  the energy request up to its $partialR$. The output of the algorithm is a bipartite graph that represents the spatio-temporal composability between the energy services and requests.\looseness=-1

\begin{algorithm}[!t]
    \footnotesize

     \renewcommand{\algorithmicrequire}{\textbf{Input:}}
    \renewcommand{\algorithmicensure}{\textbf{Output:}}
    \caption{ Partial Services Bipartite Graph Generation }
    \label{alg:PBGG}
    \begin{algorithmic}[1]
        \Require
        $\mathcal{ES},\; \mathcal{ER}, T$
        \Ensure $PBG$
         \Statex  \textbf{// Bipartite Graph Generation}     
        \State $PBG$ = Services\_Bipartite\_Graph\_Generation ($\mathcal{ES},\; \mathcal{ER}, T$)
 
            \For{$ES_i\; \textbf{\textit{in}}\; \mathcal{ES}$}
                \For{$ER_j\; \textbf{\textit{in}}\; \mathcal{ER}$}
                    \If{Nearby($ES_i.l, ER_j.l$)}

                          \Statex  \;\;\;\;\; \; \; \; \;\;\; \textbf{// Partial-Temporal Composability} 
                         \If{$ES_i.[st, et] \cap ER_j.[st, et] $}
                            \Statex \;\;\;\;\; \; \; \; \;\;\;\;\; \; \; \textbf{//Partial  Graph Generation} 
                            \State $ d = Time\_overlap (ES_i.[st, et], ER_j.[st, et])$
                          \State $partialR = d * Er_j.r/ ER_j.[st, et]$
                            \State  $PBG.$Add\_edge($ES_i.EID, ER_j.RID, partialR$)

                        \EndIf
                    \EndIf
                \EndFor
            \EndFor
        \Return $PBG$
    \end{algorithmic}
\end{algorithm}

\vspace{-10pt}
\subsubsection{\textbf{Partial Maximum Flow Matching Composition}}

In this step, we propose a \textit{PartialFlowComp}, an extension of the Maximum-Flow matching Composition proposed in Section \ref{MaxFlowSec}.  \textit{PartialFlowComp} uses the graph generated by \textit{PartialGraph} approach $PBG$ (see Sec.\ref{PBGSec}). $PBG$ will be used instead of the bipartite graph $BG$ in Algorithm \ref{alg:ESC} by changing Line 5 to call the function of Algorithm \ref{alg:PBGG}.

\vspace{-15pt}
\section{Experiments Results}
  \label{ExpSection}
  We compared our proposed composition techniques (\textit{EnergyFlowComp}) and  (\textit{PartialFlowComp}) with two resource matching algorithms, namely, first come first served algorithm (\textit{Baseline-Matching}) and Priority-based algorithm (\textit{Priority-Matching}) \cite{kruse2007data}. In \textit{Baseline-Matching}, the requests nodes in the graph were matched to services based on their start time order. For instance,  using the network of Fig.\ref{PNF_Example}(A), ES1 was assigned to ER1. In \textit{Priority-Matching}, the requests nodes in the graph were matched to services based on the size of their request, i.e., the weight of the edge. For instance,  using the network of Fig.\ref{PNF_Example}(A), ES1 was assigned to ER3. We ran a set of experiments with different settings to evaluate the \textit{effectiveness} and \textit{efficiency} of each approach.\looseness=-1 

 \begin{figure*}
    \setlength{\abovecaptionskip}{-1pt}
     \setlength{\belowcaptionskip}{-20pt}
    \centering
    \subfloat[]{\includegraphics[width=0.25\textwidth]{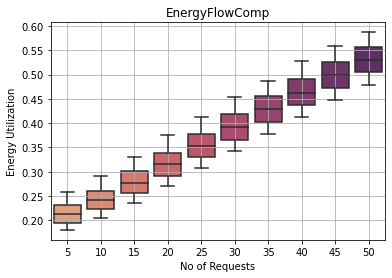}}
    \subfloat[]{\includegraphics[width=0.25\textwidth]{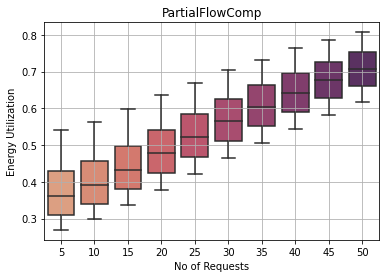}}
    \subfloat[]{\includegraphics[width=0.25\textwidth]{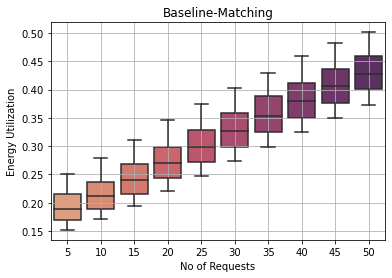}}
    \subfloat[]{\includegraphics[width=0.25\textwidth]{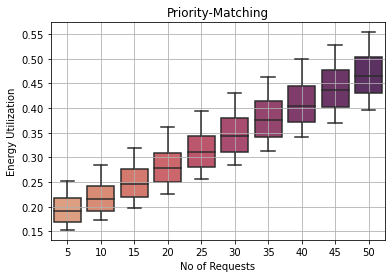}}
     \caption{Distributions of the energy utilization $EU$  score vs. number of requests }
    \label{EU_Box_R}
\end{figure*} 
\begin{figure*}
    \centering
    \setlength{\abovecaptionskip}{-1pt}
     \setlength{\belowcaptionskip}{-20pt}
    \subfloat[]{\includegraphics[width=0.25\textwidth]{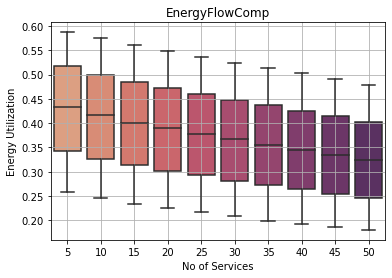}}
    \subfloat[]{\includegraphics[width=0.25\textwidth]{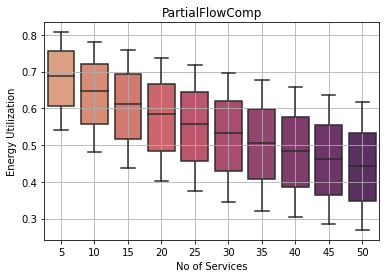}}
    \subfloat[]{\includegraphics[width=0.25\textwidth]{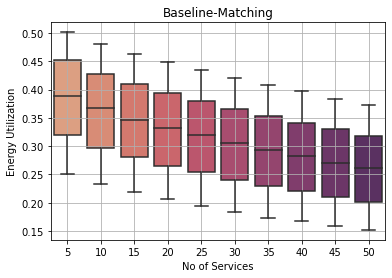}}
    \subfloat[]{\includegraphics[width=0.25\textwidth]{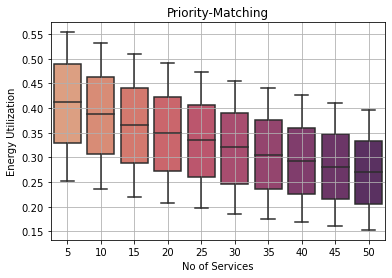}}
     \caption{Distributions of the energy utilization $EU$  score vs. number of services }
    \label{EU_Box_S}

\end{figure*} 
\begin{table}
\setlength{\abovecaptionskip}{1pt}
\setlength{\belowcaptionskip}{-8pt}
\centering
\caption{Statistics of the experiments environment}
\label{ExpVariables}

\begin{tabular}{l|l}
\hline
Variables & Value   \\ \hline
Total Energy Requests for\\ coffee shop 1 in April            & 16830   \\ 
Energy Services                  &{100000}             \\ 
Duration of All Energy Services             & {5-60 minutes} \\ 
Duration of Energy Requests      & {5-60 minutes}   \\
Provided Energy                  & {5-100 mAh}      \\
Requested Energy                 & {5-100 mAh}       \\
  \hline
\end{tabular}

\end{table}

\vspace{-10pt}
\subsection{Dataset Description}
\vspace{-5pt}

We used a real dataset generated from the developed app in \cite{yang2023Monitoring}\cite{yao2022wireless}\cite{yang2022towards}. The dataset consists of energy transfer records between a provider (smartphone) and a consumer (smartphone). The records attributes are the provider ID, consumer ID, transaction date, time, energy services' and requests' amount, and transfer duration. We use the energy dataset to generate the QoS parameters for the energy services and requests. For instance, the amount of a wireless charging transfer in mAh is used to define the amount of requested/provided energy. In addition,  the energy dataset records of a wireless charging transfer duration are used to define the end time of each request/service. \looseness=-1 

We augmented the energy sharing dataset to mimic the crowd's behavior within microcells by utilizing a dataset published by IBM for a coffee shop chain in New York City\footnote{https://ibm.co/2O7IvxJ}. The dataset consists of transaction records of customers' purchases in coffee shops for one month. Each coffee shop consists of, on average, 560 transnational records per day and 16,500 transaction record in total. We use the IBM dataset to simulate the spatio-temporal features of energy services and requests. Our experiment uses the consumer ID, transaction date, time, location, and coffee shop ID from each record in the dataset to define the spatio-temporal features of energy services and requests, e.g., start and location of energy service or a request. Tables \ref{tab:simparam} and \ref{ExpVariables} present statistics about the used datasets.\looseness=-1 
\vspace{-3pt}

\vspace{-10pt}
\subsection{Evaluation of the Composition Framework}
We ran  an extensive set of experiments to determine the \textit{effectiveness} and \textit{efficiency} of our proposed approaches. We evaluate the effectiveness of each approach by measuring the energy utilization, fulfillment ratio, consumers'  satisfaction, and providers' satisfaction. Moreover, we evaluate the efficiency of each approach by comparing their execution time while varying the number of energy services and energy requests. We run the approaches in different settings by changing the number of (1) requests and (2) services. We gradually increased each setting over the time interval $T$. We repeated the experiment 10,000 times at each point and considered the average value for each approach.\looseness=-1


\vspace{-4pt}
\subsubsection{Energy Utilization Evaluation}
\label{expr_EU}
In this subsection, we evaluate the effectiveness of the approaches in Energy Utilization ($EU$). As previously stated, the objective of the composition is to maximize the utilization of the available energy. Therefore, a high $EU$ of a composition indicates its effectiveness in utilizing the available energy. The $EU$ is computed using  Eq.\ref{Eq:EU}.

\begin{figure}[!t]
    \centering
    \setlength{\abovecaptionskip}{1pt}
    \setlength{\belowcaptionskip}{-6pt}
    \includegraphics[width=0.6\linewidth]{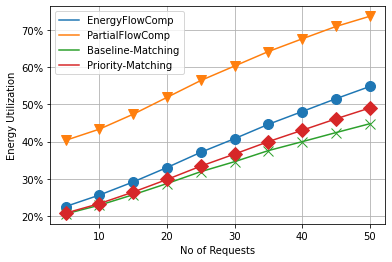}
    \caption{The average of the energy utilization $EU$ vs. number of requests}
    \label{fig:R_EU}
\end{figure}



The first experiment compared the average $EU$ of all approaches by changing the number of energy requests (see Fig.\ref{fig:R_EU}). The number of services in this experiment was a randomly selected set of 20 services with each run. In Fig.\ref{fig:R_EU}, all approaches $EU$ increase as the number of requests increases, which can be explained by the increase in the energy demand. The higher the number of requests, the higher the probability of better allocating the available energy. Furthermore, both proposed approaches perform better than the other two approaches. The \textit{EnergyFlowComp} approach performs better as it leverages the \textit{fragmented} nature of energy. Specifically, when needed, energy may be shared among requests to guarantee the maximum possible energy utilization. Similarly, the \textit{PartialFlowComp} approach outperforms all other approaches as it leverages the \textit{fragmented} nature of energy in addition to considering the \textit{partially overlapping} requests and services. Recall that considering partial requests and services will result in a more connected network flow and thereby a higher chance to \textit{efficiently} allocate the available energy.

\begin{figure}[!t]
     \centering    \setlength{\abovecaptionskip}{1pt}
    \setlength{\belowcaptionskip}{-5pt}
    \includegraphics[width=0.6\linewidth]{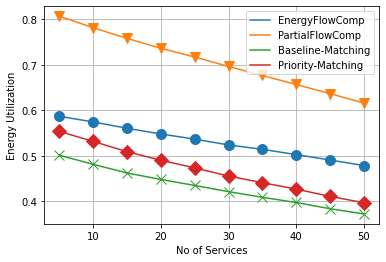}
    \caption{The average of the energy utilization $EU$ vs. number of services}
    \label{fig:S_EU}
\end{figure}

The second experiment compared the ${EU}$ by varying the number of services (see Fig.\ref{fig:S_EU}). The number of requests in this experiment was a randomly selected set of 50 requests with each run. In Fig.\ref{fig:S_EU}, the $EU$ decreases with the increased services for all the approaches. The increase in service availability can explain this observation. As services increase, the unused available energy will increase regardless of the approach. Furthermore, both proposed approaches performed better than the other two approaches for the same aforementioned reasons in the first experiment.\looseness=-1

We analyzed the distribution of the $EU$ score to ascertain the previous $EU$ evaluation. The previous experiments only considered the average $EU$ of the available services. In this set of experiments, we varied two settings: (a) the number of requests (see Fig.\ref{EU_Box_R}) and (b) the number of services (see Fig.\ref{EU_Box_S}). All approaches have a lower $EU$ dispersion with changing requests in contrast to changing services. The availability of services can justify this observation. Since $EU$ relies on the amount of available energy, changing the requests will still ensure high $EU$ as shown in Fig.\ref{fig:R_EU}. However, changing the services may lower the $EU$ as shown in Fig.\ref{fig:S_EU}.


\vspace{-8pt}
\subsubsection{Fulfillment Rate Evaluation}
\label{expr_FR}
\begin{figure}[!t]
    \setlength{\abovecaptionskip}{-3pt}
      \setlength{\belowcaptionskip}{-25pt}
    \centering
    \subfloat[]{\includegraphics[width=0.25\textwidth]{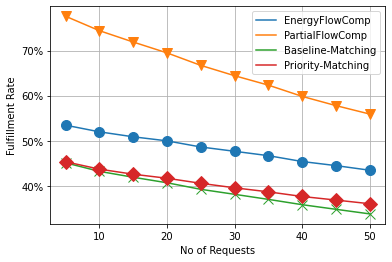}}
    \subfloat[]{\includegraphics[width=0.25\textwidth]{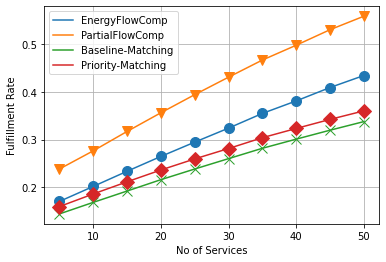}}
     \caption{The average of the fulfillment rate $FR$ vs. number of (a)  requests and (b) services}
    \label{fig:FR_R_S}
\end{figure}

We evaluated the effectiveness of the approaches in fulfilling the available energy requests. As aforementioned, better energy utilization may result in a better Fulfillment Rate ($FR$). Therefore, a higher $FR$ indicates efficiency in fulfilling the energy demand. $FR$ is computed using  Eq.\ref{Eq:FR}.\looseness=-1\\
Fig.\ref{fig:FR_R_S} (a) presents the average $FR$ for each approach.  The number of services in this experiment is a randomly selected set of 50 services with each run. In Fig.\ref{fig:FR_R_S} (a), all approaches $FR$ decrease as the number of requests increases, which can be explained by the increased amount of requested energy. The higher the number of requests with a fixed number of energy services, the lower the percentage of fulfillment. Moreover, the \textit{EnergyFlowComp} approach performed better than approaches as it leveraged the \textit{fragmented} nature of energy, as previously explained. Similarly, the \textit{PartialFlowComp} approach outperformed all approaches as it  leveraged the \textit{fragmented} nature of energy in addition to considering the \textit{partially overlapping} requests and services.\looseness=-1

The fifth experiment compared the ${FR}$ by varying the number of services (see Fig.\ref{fig:FR_R_S} (b)). The number of requests at each run is a randomly selected set of 50 requests. In Fig.\ref{fig:FR_R_S} (b), the $FR$ increases with the increase of services for all the approaches. The increase in service availability can explain this observation. As services increase, more requests will be fulfilled regardless of the approach. Furthermore, both proposed approaches performed better than the other two for the same reasons in the previous experiment.



\vspace{-10pt}
\subsubsection{Consumers' Satisfaction Evaluation}
In this subsection, we evaluate the effectiveness of the approaches in satisfying consumers. As previously stated, better energy utilization may result in  higher consumer satisfaction. We assess  Consumers' Satisfaction ($CS$) through different metrics, namely, the mean, standard deviation, and entropy satisfaction. The $CS$ is computed using  Eq.\ref{Eq:CS}, which reflects the percentage of acquired energy per request.

 \begin{figure}[!t]
    \setlength{\abovecaptionskip}{-3pt}
      \setlength{\belowcaptionskip}{-5pt}
    \centering
    \subfloat[]{\includegraphics[width=0.25\textwidth]{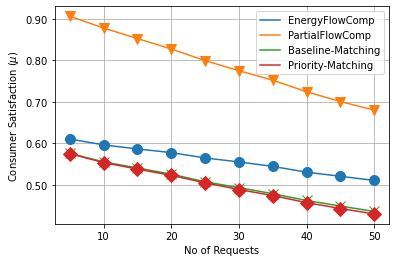}}
    \subfloat[]{\includegraphics[width=0.25\textwidth]{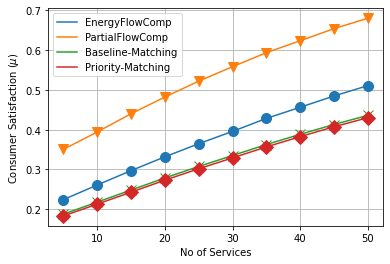}}
     \caption{The mean of consumer satisfaction $CS$ vs. number of (a)  requests and (b) services}
    \label{fig:CS_R_S}
\end{figure} 

Fig.\ref{fig:CS_R_S} (a) compares the mean $CS$ of all approaches while changing the number of requests. Intuitively, all approaches $CS$ decrease as the number of requests increases. The higher the number of requests, the lower the percentage of fulfillment and hence the lower the satisfaction. Similarly, Fig.\ref{fig:CS_R_S} (b) compares the mean $CS$ of all approaches by changing the number of services. Intuitively, all approaches $CS$ increase with the increase of services. The increase in service availability can explain this observation. As services increase, more requests will be fulfilled, which results in higher $CS$. Furthermore, both proposed approaches perform better than the other two approaches as they have a higher $FR$  as explained in Section \ref{expr_FR}.

 \begin{figure}[!t]
    \setlength{\abovecaptionskip}{-3pt}
      \setlength{\belowcaptionskip}{-25pt}
    \centering
    \subfloat[]{\includegraphics[width=0.25\textwidth]{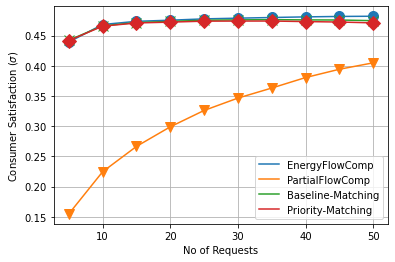}}
    \subfloat[]{\includegraphics[width=0.25\textwidth]{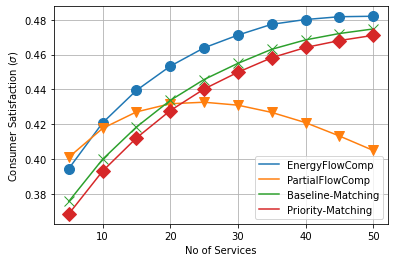}}
     \caption{The standard deviation score of consumer satisfaction $CS$ vs. number of (a)  requests and (b) services}
    \label{fig:CSS_R_S}
\end{figure}
 \begin{figure}[!t]
    \setlength{\abovecaptionskip}{-3pt}
      \setlength{\belowcaptionskip}{-5pt}
    \centering
    \subfloat[]{\includegraphics[width=0.25\textwidth]{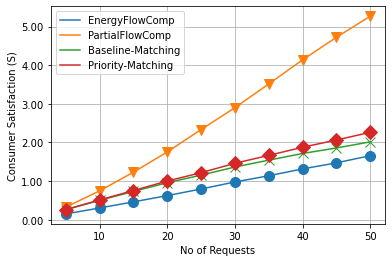}}
    \subfloat[]{\includegraphics[width=0.25\textwidth]{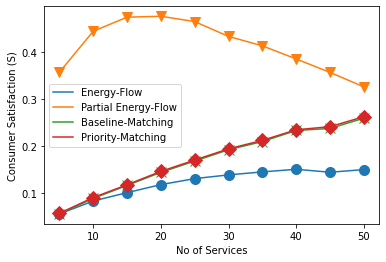}}
     \caption{The entropy score of consumer satisfaction $CS$ vs. number of (a)  requests and (b) services}
    \label{fig:CSE_R_S}
\end{figure} 
 The eighth  experiment presented the dispersion of $CS$  around the mean in both settings of changing the number of (1) requests (see Fig.\ref{fig:CSS_R_S} (a)), and (2) services  (see Fig.\ref{fig:CSS_R_S} (b)).  This metric reflects the variation of the $CS$ across requests. In Fig.\ref{fig:CSS_R_S} (a), as the number of requests increase,  the $CS$ standard deviation  converges.  This is due to the decrease of $CS$ among most of the requests as presented in Fig.\ref{fig:CS_R_S} (a). Additionally, \textit{PartialFlowComp} has the least dispersion because it offers the highest $CS$ as shown in Fig.\ref{fig:CS_R_S} (a).  In  Fig.\ref{fig:CSS_R_S} (b), we used the same $CS$ standard deviation metric with 50 random requests and varied the number of services. The figure shows that as the number of services increases, all approaches $CS$ standard deviations converged. In addition, \textit{PartialFlowComp} has the least dispersion because it offers the highest $CS$ as shown in Fig.\ref{fig:CS_R_S} (b).

The information entropy captures the multi-modal dispersion and irregularities in the distribution of the satisfaction score of consumers (see Fig.\ref{fig:CSE_R_S}). We leverage the information entropy to assess the global consumers' satisfaction (see Sec.\ref{entropyS_Sec}. A lower entropy value means a lower disorder and a higher consistency in the satisfaction score among consumers. It is worth mentioning that the entropy metric can capture small variations in the satisfaction score when the number of requests is large. These variations cannot be noticed only with the standard deviation (see Fig.\ref{fig:CSS_R_S}). We ran two experiments to assess the entropy scores of consumers' satisfactions by varying the number of  (1) requests (see Fig.\ref{fig:CSE_R_S} (a)), and (2) services (see Fig.\ref{fig:CSE_R_S} (b)). We used Eq.\ref{Eq:CSE} to measure the entropy of consumers' satisfactions.  In both experiments,  both \textit{Baseline-Matching} and \textit{Priority-Matching}  approaches have lower scores of entropy than \textit{PartialFlowComp}. This can be explained by the fact that most of the consumers' satisfaction score using both approaches is closer to zero, unlike \textit{PartialFlowComp}. Moreover,  the low score of \textit{EnergyFlowComp} approach demonstrates that not only does it offer a higher satisfaction compared to both \textit{Baseline-Matching} and \textit{Priority-Matching} but also the satisfaction score is highly consistent among consumers. In addition, \textit{PartialFlowComp} has the highest score as requests increase, but it converges when services increase. This shows that \textit{PartialFlowComp}  doesn't guarantee equal individual satisfaction for consumers when the demand is higher than the available energy. This may be addressed in future work.\looseness=-1

\vspace{-10pt}
\subsubsection{Providers' Satisfaction Evaluation}
We evaluate the effectiveness of the approaches in satisfying providers. As previously stated, a better energy utilization may result in {\em satisfying} more providers by using their services and thereby getting more rewards \cite{abusafia2020incentive}. We assess   Providers' Satisfaction ($PS$) through different metrics, namely, the mean, standard deviation, and entropy satisfaction. The $PS$ is computed using  Eq.\ref{Eq:PS}, which reflects the percentage of utilized energy per service.\looseness=-1a higher consumers satisfaction. We assess  Consumers' Satisfaction ($CS$) through different metrics: the mean, standard deviation, and entropy satisfaction. The $CS$ is computed using  Eq.\ref{Eq:CS}, which reflects the percentage of acquired energy per request.

\begin{figure}[!t]
    \setlength{\abovecaptionskip}{-3pt}
      \setlength{\belowcaptionskip}{-25pt}
    \centering
    \subfloat[]{\includegraphics[width=0.25\textwidth]{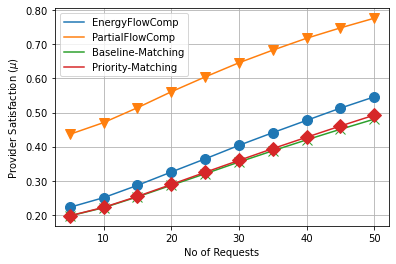}}
    \subfloat[]{\includegraphics[width=0.25\textwidth]{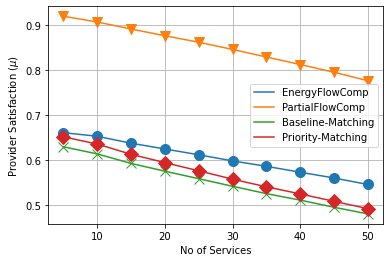}}
     \caption{The mean of providers' satisfaction $PS$ vs. number of (a)  requests and (b) services}
    \label{fig:PS_R_S}
\end{figure}

Fig.\ref{fig:PS_R_S} (a) compares the mean $PS$ of all approaches while changing the number of requests. Intuitively, all approaches $PS$ increase as the number of requests increases. This is because the higher the number of requests, the higher the probability of better allocating the available energy and, thereby, the higher the providers' satisfaction. Similarly, Fig.\ref{fig:PS_R_S} (b) compares the mean $PS$ of all approaches by changing the number of services. Intuitively, all approaches $PS$ decrease with the increase of services. The increase in service availability can explain this observation. As services increase, more services will be needed, which results in lower $PS$. Furthermore, both proposed approaches performed better than the other two approaches as they have a higher $EU$, as explained in Section \ref{expr_EU}.
\begin{figure}[!t]
    \setlength{\abovecaptionskip}{-3pt}
      \setlength{\belowcaptionskip}{-5pt}
    \centering
    \subfloat[]{\includegraphics[width=0.25\textwidth]{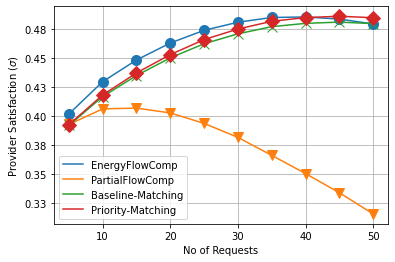}}
    \subfloat[]{\includegraphics[width=0.25\textwidth]{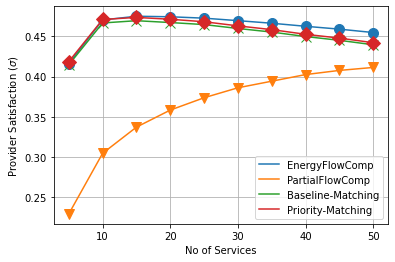}}
     \caption{The standard deviation score of providers' satisfaction $PS$ vs. number of (a)  requests and (b) services}
    \label{fig:PSS_R_S}
\end{figure} 


The following  experiment presents the dispersion of $PS$  around the mean in both settings of varying the number of (1) requests (see Fig.\ref{fig:PSS_R_S} (a)), and (2) services (see Fig.\ref{fig:PSS_R_S} (b)).  This metric reflects the variation of the $PS$ across services. In Fig.\ref{fig:PSS_R_S} (a), as the number of requests increase,  the $PS$ standard deviation's  converges.  This is due to the increase of $PS$ among most of the services as presented in Fig.\ref{fig:PS_R_S} (a). Additionally, \textit{PartialFlowComp} has the least dispersion because it offers the highest $PS$, as shown in Fig.\ref{fig:PS_R_S} (a).  In  Fig.\ref{fig:PSS_R_S} (b), we used the same metric of $PS$ standard deviation with 20 random requests and varied the number of services. The figure shows that as the number of services increases, all approaches $PS$ standard deviations converge. This is due to the decrease of $PS$ among most of the services as presented in Fig.\ref{fig:PS_R_S} (b). In addition, \textit{PartialFlowComp} has the least dispersion because it offers the highest $PS$ as shown in Fig.\ref{fig:PS_R_S} (b).\looseness=-1



 \begin{figure}[!t]
    \setlength{\abovecaptionskip}{-3pt}
      \setlength{\belowcaptionskip}{-23pt}
    \centering
    \subfloat[]{\includegraphics[width=0.25\textwidth]{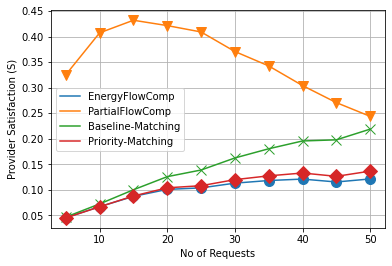}}
    \subfloat[]{\includegraphics[width=0.25\textwidth]{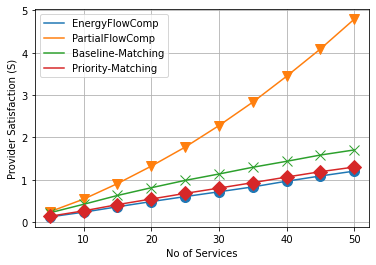}}
     \caption{The entropy score of providers' satisfaction $PS$  vs. number of (a)  requests and (b) services}
    \label{fig:PSE_R_S}
\end{figure} 
 \begin{figure}[!t]
    \setlength{\abovecaptionskip}{-3pt}
      \setlength{\belowcaptionskip}{-5pt}
    \centering
    \subfloat[]{\includegraphics[width=0.25\textwidth]{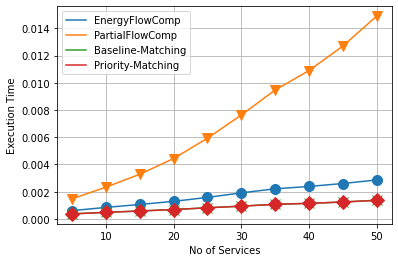}}
    \subfloat[]{\includegraphics[width=0.25\textwidth]{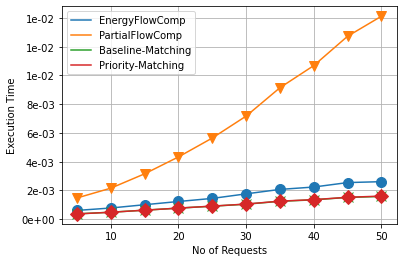}}
     \caption{The average execution time vs. number of (a)  requests and (b) services}
    \label{fig:EXE_R_S}
\end{figure}

Similar to consumers' satisfaction, we leverage the information entropy to assess the global providers' satisfaction (see Sec.\ref{entropyS_Sec}). As aforementioned, a lower entropy value means a lower disorder and a higher consistency in the satisfaction score among providers. We ran two experiments to assess the entropy score of providers' satisfactions by varying the number of  (1) requests (see Fig.\ref{fig:PSE_R_S} (a)), and (2) services (see Fig.\ref{fig:PSE_R_S} (b)). We used Eq.\ref{Eq:PSE} to measure the entropy of providers' satisfaction. In both experiments,  both \textit{Baseline-Matching} and \textit{Priority-Matching}  approaches have lower scores of entropy than \textit{PartialFlowComp}. This may be explained by the fact that most of the providers' satisfaction score using both approaches is closer to zero, unlike \textit{PartialFlowComp}. Moreover,  the low score of \textit{EnergyFlowComp} approach demonstrates that not only does it offer a higher satisfaction compared to both \textit{Baseline-Matching} and \textit{Priority-Matching} but also the satisfaction score is highly consistent among providers. Additionally, \textit{PartialFlowComp} has the highest score as services increase, but it converges when requests increase. This shows that \textit{PartialFlowComp}  doesn't guarantee an equal individual satisfaction for providers when the available services is higher than the required energy. This may be addressed in future work.



\vspace{-8pt}
\subsubsection{Computation Efficiency Evaluation}
The last set of experiments evaluates the computation cost of all approaches. Fig.\ref{fig:EXE_R_S} shows the average execution time of each approach. The execution time for all approaches increases as the number of requests/services increases. Intuitively,  the more requests/services available, the more processing is required to match these requests to services. Moreover, \textit{PartialFlowComp} has the highest computation cost as it considers services and requests that partially match. Recall that \textit{PartialFlowComp} results in a more connected network that requires more processing (see Fig.\ref{PNF_Example}).

\vspace{-15pt}
\section{Related Work}
\label{RelatedSection}


The background of our work comes from two different areas, namely, \emph{wireless energy sharing and service selection and composition}. We describe the related work in each area.\looseness=-1

\vspace{-15pt}
\subsection{ Energy Sharing}

Wireless energy charging is growing rapidly in the fields of wireless sensor networks, internet of things, mobile social networks, vehicular ad hoc networks, and UAV networks \cite{dhungana2020peer}\cite{lakhdari2020Vision}. The advent of wireless charging makes the harvested energy from IoT devices more flexible and convenient to be easily shared. Energy sharing helps create self-sustained systems. Different techniques have been developed for wireless charging in IoT, and sensor networks \cite{huang2018reconfigurable}. The most common methods are magnetic inductive coupling, magnetic resonant coupling, and microwave radiation. These techniques are used in wireless sensor networks by deploying charger robots in the network to charge the low battery sensors \cite{na2018energy}. A new paradigm of radio wave-based uncoupled wireless charging has enabled sharing and accessing harvested energy from IoT devices \cite{bell2019systems}. Multiple wireless charging techniques have been introduced in the internet of things, and wireless sensor networks \cite{huang2018reconfigurable}. For instance, a reliable energy supply method was proposed to charge low battery IoT devices using a mobile charger \cite{na2017energy}. 


Energy sharing services have been introduced as a ubiquitous alternative solution to charge IoT devices  \cite{lakhdari2020Vision}\cite{abusafia2022services}. 
Several studies have addressed challenges related to fulfilling the requirements of energy consumers \cite{lakhdari2018crowdsourcing}\cite{lakhdari2020Elastic}\cite{lakhdari2020fluid}. A temporal composition algorithm was proposed to compose energy services to fulfill a consumer's energy requirement \cite{lakhdari2018crowdsourcing}. The algorithm proposed the use of partial services and fractional knapsack to maximize the provided energy. An elastic composition was proposed to address the reliability of highly fluctuating energy providers \cite{lakhdari2020Elastic}. The composition uses the concepts of soft and hard deadlines to extend the stay of a consumer and select more reliable services. Mobility pattern impact on IoT energy services was addressed in \cite{lakhdari2020fluid}\cite{lakhdari2021proactive}. The intermittent behavior of energy services was addressed by a fluid approach \cite{lakhdari2020fluid}. The approach uses the mobility patterns of the crowd to predict the intermittent disconnections in energy services and then replace or tolerate these disconnections. Another study proposed the use of energy services as a tool for increasing consumer satisfaction \cite{abusafia2022maximizing}\cite{abusafia2022Quality}. Other studies tackled  challenges from a provider's perspective \cite{abusafia2020incentive}\cite{abusafia2020reliability}. A context-aware incentive model was proposed to address the resistance to providing energy services \cite{abusafia2020incentive}.
Another article addresses the commitment of energy consumers to receive their initiated requests \cite{abusafia2020reliability}. The paper proposes a reliability model to evaluate consumers and compose the most reliable consumers for a single energy provider. Another study proposed a model to estimate the energy loss in sharing energy services \cite{yang2023energy}. Existing literature in energy services addresses issues for a \textit{single} consumer or  provider \cite{lakhdari2020Vision}. To the best of our knowledge, challenges related to matching \textit{multiple} consumers to \textit{multiple} providers are yet to be addressed.\looseness=-1

\vspace{-15pt}
\subsection{Service Selection and Composition}

Selecting and composing services is one of the techniques in service computing. Service composition is utilized in different domains such as cloud computing \cite{zeng2004qos}, social networks \cite{maaradji2011social}, and the Internet of Things \cite{sun2019energy}\cite{zeng2020towards}. Services are composed according to their \textit{functional properties} and consumer preferences (QoS). Most service composition methods convert the composition into a resource scheduling or an optimization problem. Resource scheduling in service composition has been extensively researched \cite{li2018service}. The fundamental parameters of resource scheduling algorithms are the target to optimize, e.g., computing or storage resources in cloud computing, and the scheduling priority policy based on time. For example, the priority for Short Job First (SJF) scheduling algorithm is the shortest jobs to be scheduled first \cite{ghanbari2012priority}. Service composition is mostly driven by resource utilization maximization or minimization of scheduling time \cite{yau2009adaptive}\cite{bae2007fairness}. Different optimization algorithms have been utilized for service composition, such as integer programming, genetic algorithm (GA), and particle swarm optimization (PSO) \cite{yau2009adaptive}\cite{bae2007fairness}. 

The composition of crowdsourced services  should consider two aspects, the spatio-temporal features of the consumers and their preferences \cite{lakhdari2020composing}. Usually, crowdsourced services are from different sources (mobile and static devices). The functional properties of the provided services should conform to the spatio-temporal features of the query. Consumer preferences have to be met by the QoS of the provided services. Crowdsourcing energy as a service is converted to a Qos-aware service composition problem \cite{lakhdari2018crowdsourcing}\cite{lakhdari2020Vision}. The spatio-temporal features of energy services are considered QoS attributes (i.e., start and end time, duration, and location of an energy service). Composing energy services relies on finding the optimal selection of nearby services that fulfill a consumer's required energy within their query duration. Existing composition techniques may not be  applicable to compose energy services due to the uniqueness of the crowdsourced IoT environment. For example, an IoT user may consume only a part of the advertised energy service.\looseness=-1


\vspace{-15pt}
\section{Conclusion}
\label{ConclusionSection}
We proposed a novel  spatio-temporal service composition framework for sharing IoT energy services among energy consumers. We designed a service model for IoT energy services that aggregate the wearable-based harvested energy per energy proxy. The aggregated energy will be stored in an energy container (i.e., the battery of a smartphone). Energy services are delivered to consumers through wireless power transfer channels. The proposed framework aims at fulfilling the requirements of multiple energy requests. We reformulated the problem of spatio-temporal service composition to provision multiple energy requests as a matching problem. We designed and developed two composition approaches that efficiently provision multiple IoT energy services to accommodate multiple energy requests. We conduct a set of extensive experiments to assess the effectiveness and efficiency of the proposed framework. Managing the energy storage limitation and the dynamic behavior of energy usage will be the focus of our future work.\looseness=-1
\vspace{-10pt}

\section*{Acknowledgment} 
This research was partly made possible by LE220100078 and DP220101823 grants from the Australian Research Council. The statements made herein are solely the responsibility of the authors.
\vspace{-10pt}
\bibliographystyle{IEEEtran}
\bibliography{ref}
\vskip -35pt plus -1fil
\begin{IEEEbiography}[{\includegraphics[width=1in,height=1.25in,clip,keepaspectratio]{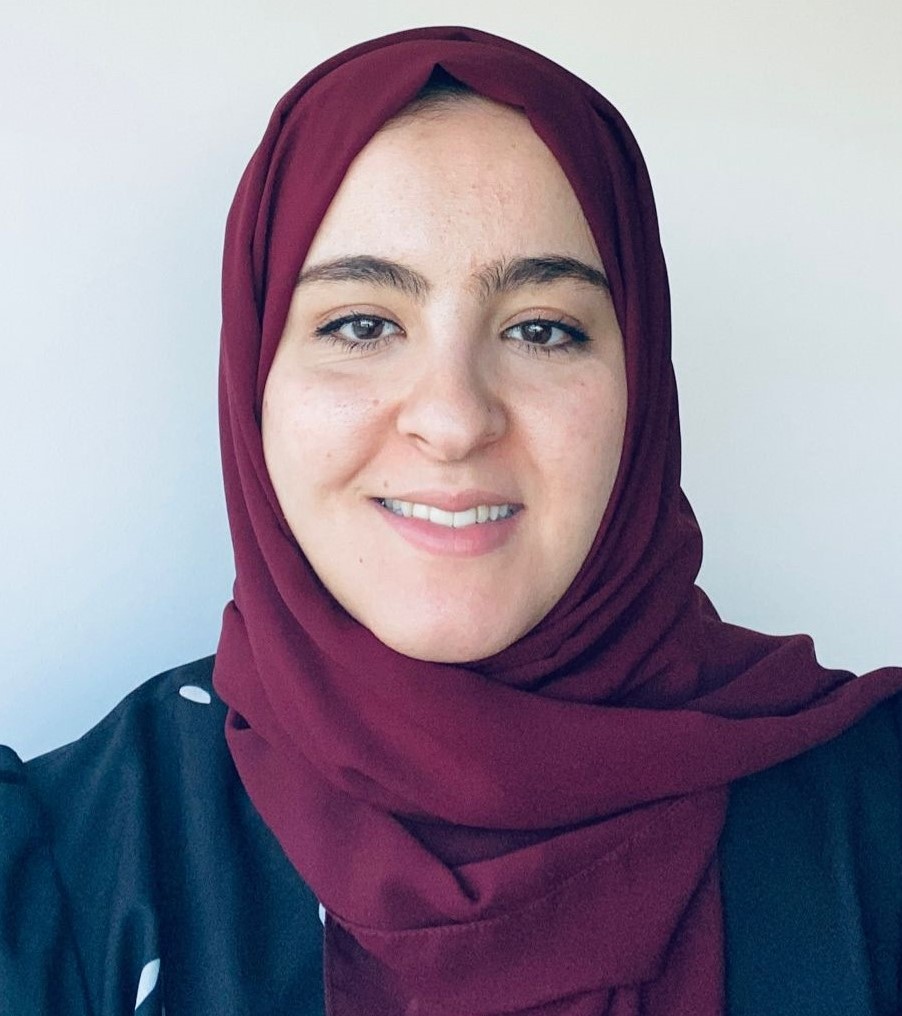}}]%
{Amani Abusafia} is a Postdoctoral Fellow in the School of Computer Science at the University of Sydney. She received her Bachelor's degree (2009) and Master's degree (2013) in Computer Science from The University of Sharjah, United Arab Emirates. She worked as a lecturer at the Department of Computer Science at the University of Sharjah for 6 years. Her research interests include Service Computing, Crowdsourcing, and IoT.\looseness=-1

\end{IEEEbiography}
\vskip -40pt plus -1fil
\begin{IEEEbiography}
[{\includegraphics[width=1.1 in,height=1.25in,clip,keepaspectratio]{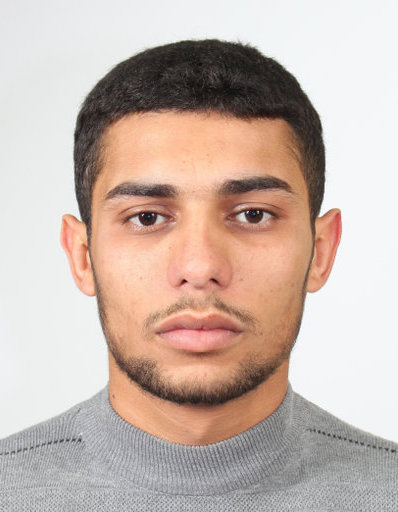}}]%
{Abdallah Lakhdari}
is a Postdoctoral Fellow in the School of Computer Science at the University of Sydney. He received his Ph.D. in Engineering from the University of Sydney, Australia. He received his Bachelor's degree (2010) and Master's degree (2013) in Computer Science from The University of Laghouat, Algeria.  He was a visiting scholar a New Mexico Tech. He worked as a lecturer at the Department of Computer Science at The University of Laghouat, Algeria. His research interests   include Social Computing, Crowdsourcing, IoT, and Service Computing.\looseness=-1
\end{IEEEbiography}
\vskip -30pt plus -1fil
\begin{IEEEbiography}[{\includegraphics[width=1.1in,height=1.25in,clip,keepaspectratio]{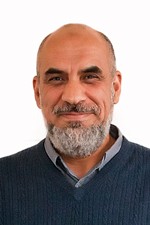}}]%
{Athman Bouguettaya}
is a Professor in the School of Computer Science at the University of Sydney. He received his Ph.D. in Computer Science from the University of Colorado at Boulder (USA) in 1992. He is or has been on the editorial boards of several journals, including the IEEE Transactions on Services Computing, ACM Transactions on Internet Technology, the International Journal on Next Generation Computing, and VLDB Journal. He is a Fellow of the IEEE and a Distinguished Scientist of the ACM. He is a member of the Academia Europaea (MAE).\looseness=-1
\end{IEEEbiography}

\end{document}